\documentclass[%
 reprint,
nofootinbib,
 amsmath,amssymb,
 aps,
]{revtex4-1}
\usepackage[english]{babel}

\newcommand{\araa}{ARA\&A}		
\newcommand{\apjl}{ApJLett}		
\newcommand{\mnras}{MNRAS}		

\usepackage{amsmath,amsfonts,amssymb}
\usepackage{mathrsfs}
\usepackage{graphicx}
\usepackage{hyperref} 
\usepackage{color}
\usepackage{bbold}
\usepackage{xfrac}
\usepackage{braket}
\usepackage{tabularx}
\usepackage{xtab,afterpage,longtable}
\usepackage{booktabs,chemformula}
\usepackage{array}
\usepackage{bm}
\newcolumntype{P}[1]{>{\centering\arraybackslash}p{#1}}
\newcolumntype{L}[1]{>{\centering\arraybackslash}l{#1}}
\usepackage{tabularx,multirow,booktabs,blindtext}
\usepackage{graphicx}
\usepackage{dcolumn}
\usepackage[mathscr]{eucal}
\newlength\replength
\usepackage{ltablex}
\usepackage{dsfont}

\newcommand\repfrac{.33}

\newcommand\rulewidth{.6pt}
\newcommand\tdashfill[1][\repfrac]{\cleaders\hbox to \replength{%
  \smash{\rule[\arraystretch\ht\strutbox]{\repfrac\replength}{\rulewidth}}}\hfill}

\newcommand\tdotfill[1][\repfrac]{\cleaders\hbox to \replength{%
  \smash{\raisebox{\arraystretch\dimexpr\ht\strutbox-.1ex\relax}{.}}}\hfill}

\newcolumntype{s}{>{\centering\arraybackslash}>{\hsize=1.25\hsize}X}
\newcolumntype{j}{>{\centering\arraybackslash}>{\hsize=7.5\hsize}X}
\newcolumntype{g}{>{\centering\arraybackslash}>{\hsize=7.65\hsize}X}
\newcolumntype{b}{>{\centering\arraybackslash}>{\hsize=7.65\hsize}X}

\begin{document}

\preprint{APS/123-QED}

\title{Ultralight Bosonic Field Mass Bounds from Astrophysical Black Hole Spin}

\author{Matthew J. Stott}
 \affiliation{Theoretical Particle Physics and Cosmology Group, Department of Physics, King’s College London, University of London, Strand, London, WC2R 2LS, United Kingdom}
 \email{Matthew.stott@kcl.ac.uk}

\date{\today}

\begin{abstract}
Black Hole measurements have grown significantly in the new age of gravitation wave astronomy from LIGO observations of binary black hole mergers. As yet unobserved massive ultralight bosonic fields represent one of the most exciting features of Standard Model extensions, capable of providing solutions to numerous paradigmatic issues in particle physics and cosmology. In this work we explore bounds from spinning astrophysical black holes and their angular momentum energy transfer to bosonic condensates which can form surrounding the black hole via superradiant instabilities. Using recent analytical results we perform a simplified analysis with a generous ensemble of black hole parameter measurements where we find superradiance very generally excludes bosonic fields in the mass ranges; Spin-0: $\{ 3.8\times10^{-14}\ {\rm eV} \leq \mu_0 \leq 3.4\times10^{-11}\ {\rm eV}, 5.5\times10^{-20}\ {\rm eV} \leq \mu_0 \leq 1.3\times10^{-16}\ {\rm eV}, 2.5\times10^{-21}\ {\rm eV} \leq \mu_0 \leq 1.2\times10^{-20}\ {\rm eV}\}$, Spin-1: $\{ 6.2\times10^{-15}\ {\rm eV} \leq \mu_1 \leq 3.9\times10^{-11}\ {\rm eV}, 2.8\times10^{-22}\ {\rm eV} \leq \mu_1 \leq 1.9\times10^{-16}\ {\rm eV} \}$ and Spin-2:  $\{ 2.2\times10^{-14}\ {\rm eV} \leq \mu_2 \leq 2.8\times10^{-11}\ {\rm eV}, 1.8\times10^{-20}\ {\rm eV} \leq \mu_2 \leq 1.8\times10^{-16}\ {\rm eV}, 6.4\times10^{-22}\ {\rm eV} \leq \mu_2 \leq 7.7\times10^{-21}\ {\rm eV} \}$ respectively. We also explore these bounds in the context of specific phenomenological models, specifically the QCD axion, M-theory models and fuzzy dark matter sitting at the edges of current limits. In particular we include recent measurements of event GW190521 and M87* used to constrain both the masses and decay constants of axion like fields. Finally we comment a simple example of a spectrum of fields for the spin-0 and spin-1 cases.

\end{abstract}

\maketitle


\section{Introduction}

Black holes (BHs) as solutions to Einstein’s field equations offer a vital probe into the fundamental interactions and potential constituents of theories that determine the nature of our Universe \cite{Hawking:1969sw,Hawking:1974sw,Hawking:1974rv,Gibbons:1977mu,Hawking:1976de,Penrose:1964wq}. Populations of BHs are often classified into mass bounds based on observational evidence to date. Common designations across the BH mass spectrum normalised to the solar mass unit $M_{\odot}$, are: stellar mass BHs \cite{Woosley:1993wj,Heger:2002by} ($5 M_{\rm \odot}\lesssim M_{\rm BH} \lesssim 10^2 M_{\rm \odot}$), intermediate mass BHs \cite{Madau:2001sc,Miller:2003sc,Miller:2001ez} (IMBHs: $10^{2} M_{\rm \odot}\lesssim M_{\rm BH} \lesssim 10^5 M_{\rm \odot}$) , low mass BHs (LMBHs) \cite{2007ApJ...670...92G,Gultekin:2014vma} ($10^{5} M_{\rm \odot}\lesssim M_{\rm BH} \lesssim 10^{6} M_{\rm \odot}$), supermassive BHs \cite{Ferrarese:2000se,Marconi:2003tg} (SMBHs: $10^{6} M_{\rm \odot}\lesssim M_{\rm BH} \lesssim 10^{10} M_{\rm \odot}$) and ultramassive BHs \cite{Shemmer:2004ph,HlavacekLarrondo:2012js} (UMBHs: $10^{10}  M_{\rm \odot}\lesssim M_{\rm BH} \lesssim 10^{12} M_{\rm \odot}$).  To date numerous BHs have been documented in the Milky Way and nearby galaxies from observed X-ray binaries \cite{Farr:2010tu}.  
Some decades after their discovery it is now also well understood most galaxies or active galactic nuclei (AGN) \cite{1995ARA&A..33..581K} harbour a SMBH at their core. Phenomenologically BHs themselves may provide interesting candidates to observational issues, such as primordial BHs representing the total dominant non-baryonic matter in the Universe \cite{Carr:2016drx,Katz:2018zrn,Montero-Camacho:2019jte}. 

\begin{figure}[t]
    \includegraphics[width=0.49\textwidth]{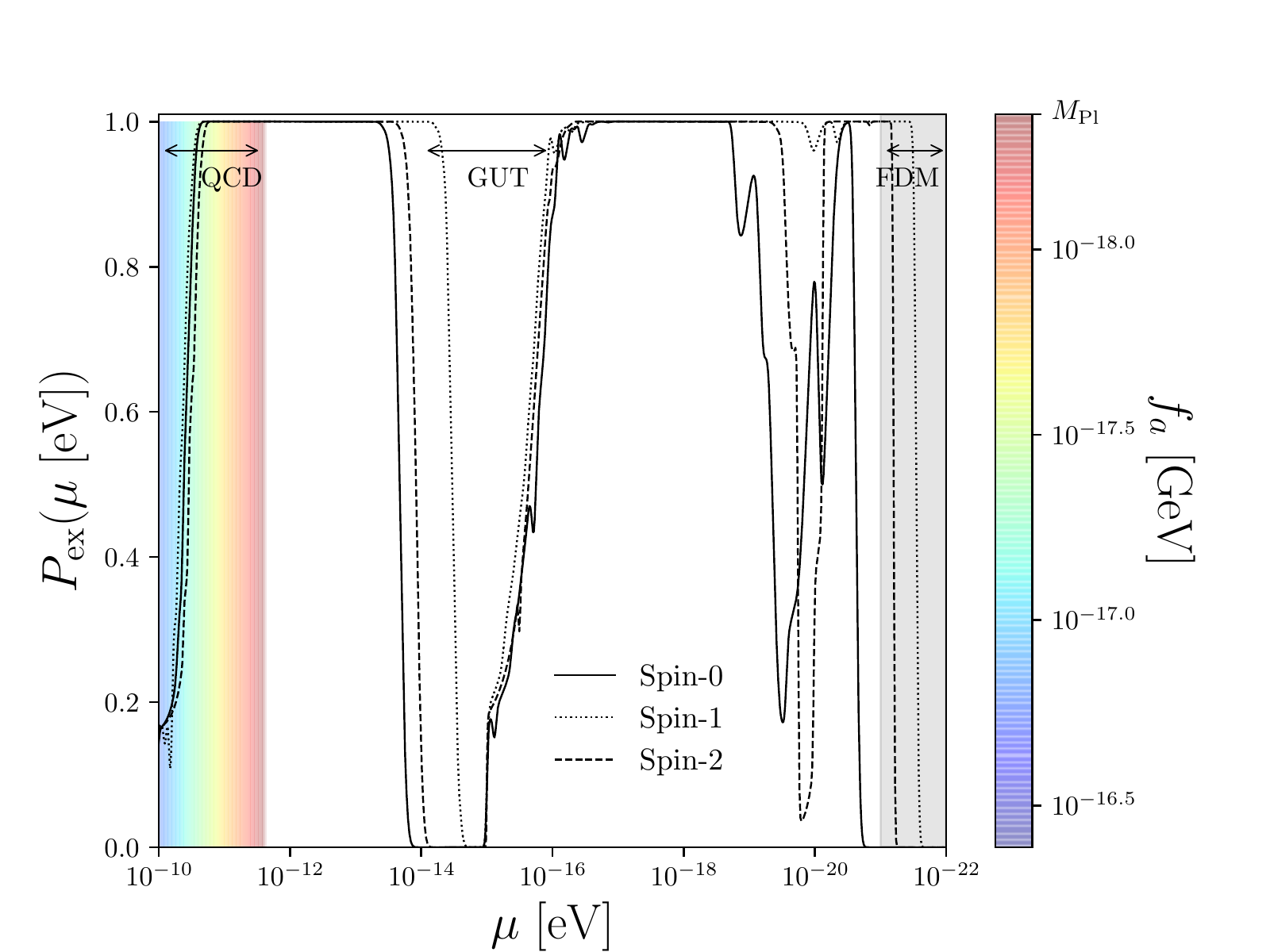}
    \caption{Total probability of exclusion functions for massive ultralight bosonic fields of spin-0 (\emph{solid line}), spin-1 (\emph{dotted line}) and spin-2 (\emph{dashed line}) from the collective BH data ensemble in Appendix~\ref{app:bhdata} ,using the measurements in Table~\ref{tab:stellarBHs}, Table~\ref{tab:ligodata} and Table~\ref{tab:SMBH}. The \emph{shaded grey} region represents expected mass limits of FDM ($10^{-(21-22)}\ {\rm eV}$)(see Section~\ref{sec:dmbounds}). The \emph{coloured shaded} region represents the bounds on the QCD axion in the limit $f_a \leq M_{\rm Pl}$, its mass/decay constant defined according to Ref.~\cite{diCortona:2015ldu}. The dip in the exclusions function represents an absence of measurements for IMBHs which for spin-0 bosons is associated to the GUT axion mass window defined in Eq.~(\ref{eq:gutaxion}).}
    \label{fig:p_explot}
\end{figure}
Recent successful probes of BHs have also taken place in the new era of \emph{gravitational wave} (GW) observational astronomy, ushered in by their historic first direct detection from the coalescence of a binary BH (BBH) system by the Laser Interferometer Gravitational-Wave Observatory (LIGO)/ VIRGO collaboration \cite{Abbott:2016blz,TheLIGOScientific:2014jea}. This success has continued with dozens of new observations  \cite{TheLIGOScientific:2016pea}, including neutron star (NS) binaries \cite{TheLIGOScientific:2017qsa,Monitor:2017mdv,GBM:2017lvd} and possible BH-NS binaries \cite{LIGOScientific:2020stg}. Recently the first direct detection of an IMBH \cite{Abbott:2020tfl,Abbott:2020mjq} ($M_{\rm BH} = 142^{+28}_{-16}M_{\odot}$) from LIGO event GW190521 appears to have involved two merger objects heavier than previously expected limits from models of supernova dynamics. The apparent shadow from the event horizon of the SMBH M87* was also recently presented by the Event Horizon Telescope (EHT) collaboration \cite{Akiyama:2019cqa,Akiyama:2019brx,Akiyama:2019bqs,Akiyama:2019eap,Akiyama:2019fyp}. This work representing the first direct experimental evidence of a BH. Collectively these landmarks indicate the initial footings of a wealth of data to come. This exciting prospect potentially capable of deepening our understanding of pertinent questions in both the broader pictures of astrophysics and particle physics, where objects on the largest scales might enlighten us about physics operating on some of the smallest.

Ultralight bosonic matter with weak couplings to the standard matter content of the Universe are common features of many grand unified theories (GUTs) \cite{Frieman:1995pm,Abel:2008ai,Essig:2013lka}. Common place in these theoretical frameworks are extended sectors of massive scalar (spin-0), vector (spin-1) and tensor (spin-2) quantum fields, which prove challenging to probe experimentally due to their highly suppressed couplings. A archetypal example is the QCD axion  \cite{Kim:1986ax,Preskill:1982cy,Weinberg:1977ma}, a pseudoscalar boson proposed to provide a dynamical field solution to the issue of CP violation in the Standard Model, via the Peccei-Quinn (PQ) mechanism \cite{Peccei:1977hh,Peccei:2006as}. The mass of this field is light, of the order, $\mu^{\rm QCD}_0 \simeq 5.71\times 10^{-6}{\rm eV} \left(\sfrac{10^{12}{\rm GeV}}{f_a}\right)$ \cite{diCortona:2015ldu}, where $f_a$ is the mass-scale where the anomalous global symmetry the field is charged under, is broken. String and M-theory compactification models often predict a plethoric landscape (string axiverse \cite{Arvanitaki:2009fg,Cicoli:2012sz,Acharya:2010zx}) of ultralight pseudoscalars or axion-like particles (ALPs), the dynamical scales of these ubiquitous degrees of freedom potentially spanning many decades\footnote{See for example Refs.~\cite{Conlon:2006tq,Choi:2006za,Higaki:2011me} for counter arguments surrounding tachyonic saxion masses and AdS vacua, where the nature of the supersymmetric locus may suggest stabilised moduli and unfixed axions forming an axiverse are non-trivial solutions.}. These fields may offer solutions to a number of core issues in cosmology through their symmetry properties \cite{Marsh:2015xka,Wantz:2009it,Kawasaki:2013ae,Preskill:1982cy}, such as cold dark matter (DM) \cite{Hu:2000ke,Hui:2016ltb,Irsic:2017yje}, quintessence/ dark energy (DE) \cite{Svrcek:2006hf,Kamionkowski:2014zda,Copeland:2006wr}, solutions to the electroweak hierarchy problem \cite{Graham:2015cka} and cosmic inflation \cite{Linde:1991km,Dimopoulos:2005ac,Kim:2004rp,Hertzberg:2008wr,Baumann:2009ds} etc. 

Extended sectors of spin-1 vector fields \cite{Goodsell:2009xc,Lees:2017lec,Curtin:2014cca} or \emph{dark photons} come from motivations for new abelian $U(1)$ gauge bosons in a hidden sector, possessing highly suppressed couplings to electrically charged particles induced through kinetic mixing with the photon. These often arise in numerous extensions of the standard model such as supersymmetric theories \cite{Abazov:2009hn} or general hidden portal models \cite{Kaneta:2016wvf}. Like string theoretic axions it is expected they could be ultralight, leading to observational effects from a cosmological standpoint, relevant to areas such as inflation \cite{Ford:1989me,Burd:1991ew}, DE \cite{Koivisto:2008xf,Heisenberg:2014rta,Boehmer:2007qa,Kiselev:2004py} or DM \cite{Feng:2009mn,ArkaniHamed:2008qn} (weakly interacting slim particles (WISPS) \cite{Arias:2012az}) etc.

Theories of massive spin-2 fields \cite{Fierz:1939ix,Hinterbichler:2012cn,Buchbinder:1999be,Folkerts:2013mra,Higuchi:1986py,Tamanini:2013xia} have traditionally faced complications such as the presence of the unstable Boulware-Deser ghost \cite{Boulware:1973my}, which is also found in non-linear extensions of massive gravity. Example solutions have been formulated  \cite{deRham:2010ik,deRham:2010kj}, such as bimetric gravity theories \cite{Hassan:2011tf,Hassan:2011zd,Hassan:2011ea,Volkov:2011an}. These ghost-free theories allow for an understanding of massive spin-2 fields coupled to gravity (i.e. multi-vielbeins \cite{Hinterbichler:2012cn} or multi-metrics \cite{Hassan:2012wt} approaches), though the specifics are certainly non-trivial to ensure no significant deviations from general relativity etc. Recent progress on the stability on particular families of bimetric gravity theories has opened up the potential of exploring the dynamics of massive spin-2 particles \cite{Brito:2015oca,Brito:2013yxa,deRham:2010ik,Hinterbichler:2011tt,deRham:2014zqa,Schmidt-May:2015vnx} via unique non-linear solutions \cite{deRham:2010kj,Hassan:2011hr,Hassan:2011zd} coupling two tensors which interact via non-derivative terms. Under fixed conditions this scenario can be considered as traditional gravity theory with an extended sector of spin-2 fields \cite{Brito:2016peb,PhysRevLett.124.211101}. It is also possible to extend these investigations to theories of massive gravity in order to explore properties of the graviton \cite{Brito:2013wya,Brito:2015oca}.

If we are to seriously consider the low energy effective limits of extensions to the Standard Model such as string theoretic frameworks, then it is natural to ask what constraints can we place on these extended sectors of ulralight bosonic fields. Fortunately a novel solution to this problem has been highlighted recently, stemming from the spacetime dynamics of rotating astrophysical BHs and the universal nature of the gravitational couplings ultralight bosons possess \cite{Arvanitaki:2010sy,Brito:2015oca,Dolan:2007mj,Brito:2013wya,Pani:2012vp,Herdeiro:2016tmi}. A rotating astrophysical BH in a four-dimensional asymptotically flat spacetime is described by the Kerr geometry \cite{Kerr:1963ud}, parameterised according to its mass, $M_{\rm BH}$ and angular momentum, $J=aM_{\rm BH}$. The phenomena of superradiance from rotational bodies is a general feature which can occur with systems that present a dissipating surface with a non-vanishing angular moment component, leading to many fascinating outcomes \cite{Brito:2015oca,Cardoso:2012zn}. The presence of perturbations close to a rotating BH are amplified \cite{Damour:1976kh,Zouros:1979iw} when enclosed in a suitable reflecting cavity. This instability is known as the \emph{BH bomb} scenario \cite{Press:1972zz}. Assuming on-shell production of yet unseen bosons from quantum fluctuations of the vacuum, naturally a suitable confining mechanism is supplied from the mass of bosonic fields leading to such an instability. The so-called \emph{no-hair} and \emph{uniqueness} theorems permit precision cosmology of BHs, inturn allowing us to probe the existence of quasi-equilibrium configurations consisting of a classical condensate or \emph{bosonic cloud} surrounding the BH. Observations of BHs with a non-zero spin can act as effective particle detectors \cite{Brito:2014wla}, offering both the possibility of
direct detection of their presence or excluding their existence, which is the focus of this work.   
 
Throughout we will refer to each test field as general ultralight boson parameterised by their intrinsic spin, using the subscript notation $\mu_s$ where $s=0$ for massive scalars, $s=1$ for massive vectors and $s=2$ for massive tensors. Both spin-0 and spin-1 bosons constraints can be considered as limits of general ALP and dark photon bounds where we also use the notation, $\mu_{\rm ax}$ and $\mu_{\rm vec}$ interchangeably with $\mu_{0}$ and $\mu_{1}$ \cite{Brito:2015oca,Cardoso:2018tly}. In principle constraints on these fields in the case of spin-1 and spin-2 fields can be extended to stringent limits on the mass of the photon \cite{Pani:2012vp} and the graviton \cite{Brito:2013wya} under certain considerations. The paper is structured as follows. In Section~\ref{sec:superradiance} we introduce the basic features of the superradiant instability and review analytical results in both the free-field limit for spin-0, -1 and -2 fields as well as the interacting limit for the spin-0 case. In Section~\ref{sec:bounds} we give our results for the bounds on ultralight bosons of spin-0, -1 and -2 using observational measurements the astrophysical BHs documented in Appendix~\ref{app:bhdata}. We also give limits on the self-interaction strength of spin-0 fields or ALPs using considerations for the possible dynamical collapse of the boson cloud (see Section~\ref{sec:bosenova}) when the coupling is sufficiently strong to alter the superradiance process. In Section~\ref{sec:massgaps} we discuss the current edges of the bounds presented in Section~\ref{sec:bounds}, in the context of specific phenomenological models. We also speculate on possible bounds from recent observations of massive compact objects where either the nature of the object or the spin is not-well defined, but the observed mass extends the results in Section~\ref{sec:bounds}. In Section~\ref{sec:multifield} we comment on a heuristical example of constraints with multiple fields utilising a spectrum of fields determined by a random matrix theory (RMT) approach, in the limit of suppressed self-interactions. Finally we conclude our discussions in Section~\ref{sec:conclusion}.

\section{Black Hole Superradiance and Ultralight Bosnic Fields}
\label{sec:superradiance}
The Kerr geometry associated to a rotating astrophysical BH in canonical Boyer-Lindquist coordinates is defined by the line element,
\begin{align}
    &ds^2_{\rm Kerr} = -\left(1-\frac{2r_gr}{\Sigma}\right)dt^2 - \frac{4r_g ar\sin^2 \theta}{\Sigma}dtd\phi \\ &+ \frac{\Sigma}{\nabla}dr^2 + \Sigma d\theta^2 + \frac{(r^2+a^2)^2-a^2\nabla \sin^2\theta}{\Sigma}\sin^2\theta d\phi^2 \ ,
    \label{eq:kerr}
\end{align}
with metric functions, $\Sigma = r^2 + a^2\cos^2 \theta$ and $r_{\pm} = r_g \pm \sqrt{r^2_g-a^2}$. The outer physical BH event horizon is the null hypersurface denoted by $r=r_+$. The Kerr spacetime omits an \emph{ergoregion} surrounding the physical horizon where time-translation Killing vectors $\partial_t$, becomes spacelike. This allows for the Killing energy density of a bosonic field to become negative under certain conditions (Eq.~(\ref{eq:condition})) leading to an effective negative Killing energy flux towards the horizon. This observed growth of the bosonic field (from an observer at spacial infinity) from the extraction of the BHs angular momentum is the \emph{superradint instability}. The efficiency of the instability is determined by the gravitational fine-structure constant, regulated by the two-key length scales of the  boson-BH system, 
\begin{equation}
    \alpha \equiv \frac{r_g}{\lambda_c}  = \frac{G_{\rm N}M_{\rm BH}\mu}{\hbar c}\ ,
\end{equation}
where $r_g \equiv \sfrac{GM_{\rm BH}}{c^2}$ defines the relevant length scale of the BH. The value of $\lambda_c=\sfrac{\hbar}{\mu_s c}$ represents the reduced Compton wavelength of a massive bosonic field. Superradiance is maximal when the Compton wave-length of the boson is comparable to the Schwarzschild radius of the BH, $\alpha \sim 1$. The exponential amplification of the bosonic field occurs until the superradiance condition is no longer satisfied,  
\begin{equation}
    0< \omega \approx \mu < m \Omega_{\rm H}\ ,
    \label{eq:condition}
\end{equation}
where $\Omega_{\rm H}$ is the angular phase velocity of the BH event horizon. This can be expressed in terms of the dimensionless spin or Kerr parameter of the spinning BH, $a_* \equiv \sfrac{a}{r_g}\in(0,1]$\footnote{A general astrophysical upper limit for this parameter is $a_* \leq 0.998$, from modelling of thin disks, known as the Thorne limit \cite{1974ApJ...191..507T}. Upper bounds generally depend heavily on the nature of the accretion flow so we fix the upper limit to 1, from the cosmic censorship conjecture. This is the value required in order for a rotating BH to possess an event horizon, vital to the superradiance process. Measured BHs which come close to saturating this bound are sometimes referred to as \emph{extremal}. }, where,  
\begin{equation}
    \Omega_{\rm H} = \frac{1}{2r_g}\frac{a_*}{1+\sqrt{1-a_*^2}}\ .
\end{equation}
In the limit of \emph{small coupling} or the \emph{non-relativistic} limit ($\alpha \ll 1$) superradiance can be considered in the perturbative regime through an expansion in powers of $\alpha$, which is the phenomenologically relevant domain for many models of massive bosons. The eigenmode spectrum and rates for a series states are denoted by the harmonic decomposition of the principle ($n$)\footnote{The principle quantum number is sometimes denoted as $\bar{n}=n+l+1$ where $n$ is the radial quantum number. Throughout our we use the notation $n$ to represent the principle quantum number only.}, orbital angular momentum ($l$), total angular momentum ($j$) and azimuthal angular momentum ($m$) quantum numbers, replicating that of the hydrogen atom. In the case of massive scalar fields we adopt the $\ket{nlm}$ notation and likewise $\ket{nljm}$ for the case of massive vectors and tensors. 
\begin{figure*}[!tbp]
  \centering
  \begin{minipage}[b]{0.49\textwidth}
    \includegraphics[width=\textwidth]{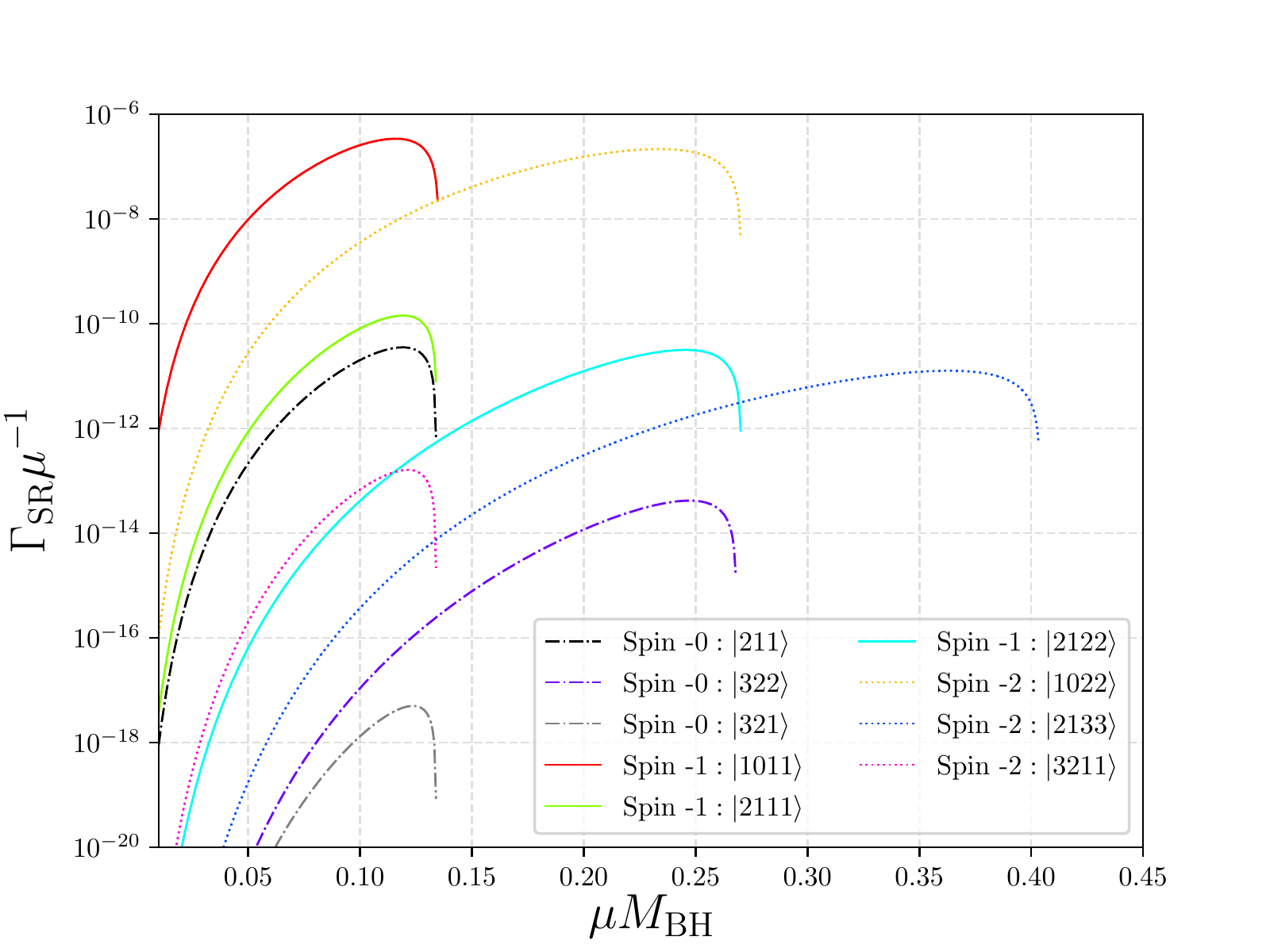}
  \end{minipage}
  \hfill
  \begin{minipage}[b]{0.49\textwidth}
    \includegraphics[width=\textwidth]{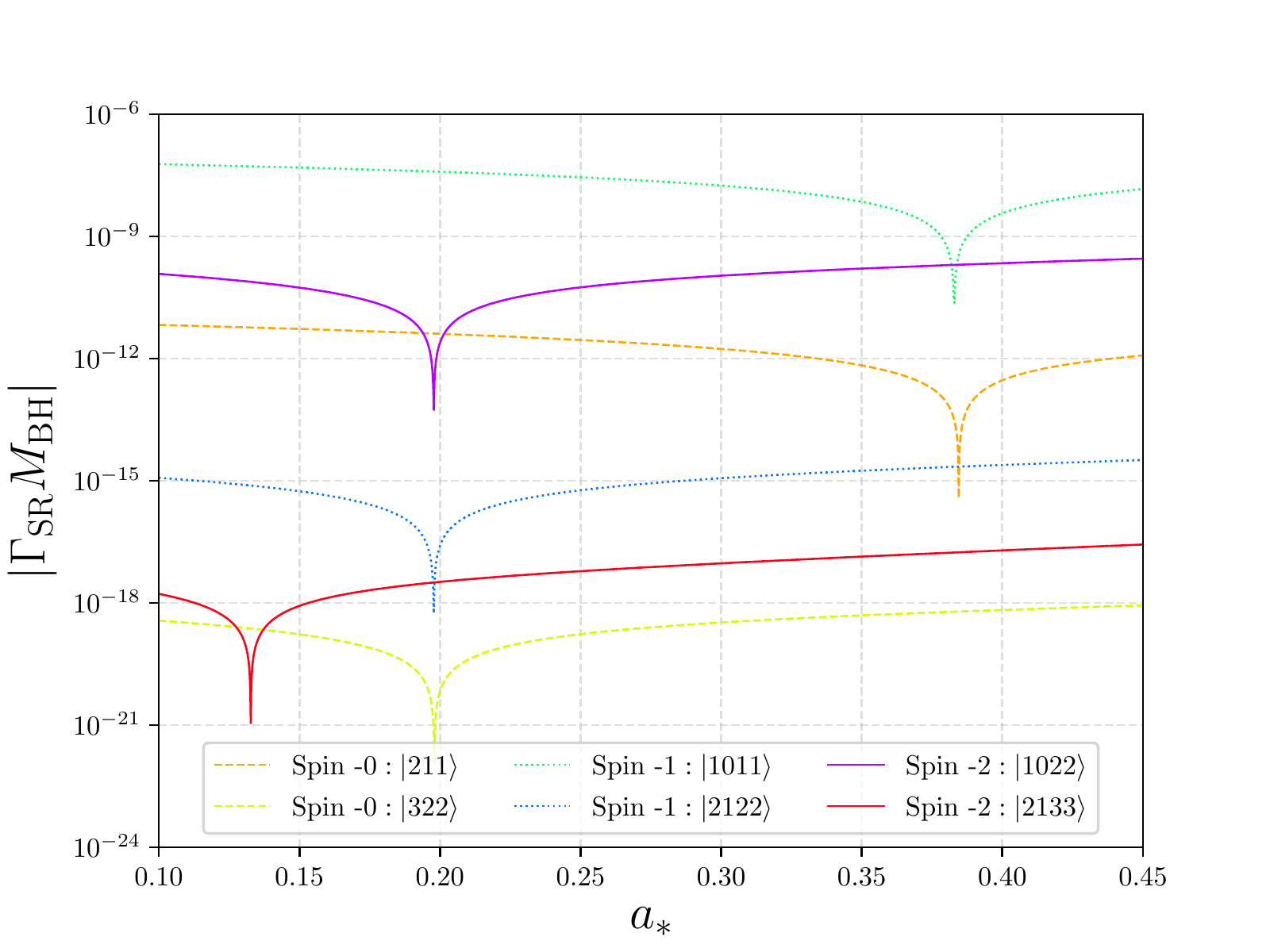}
  \end{minipage}
  \caption{Imaginary component of the bound-state frequency $\omega$ in Eq.~(\ref{eq:supstates}), representing the instability rate, denoted as either $\Gamma_{nlm}$ or $\Gamma_{nljm}$. For spin-0 ($\ket{nlm}$) or spin-1 states ($\ket{nljm}$) we use the analytical results found in Refs.~\cite{PhysRevD.22.2323,Baumann:2019eav,Baryakhtar:2017ngi}.  For spin-2 ($\ket{nljm}$) states we use the analytical results of Ref.~\cite{PhysRevLett.124.211101}. In the \emph{left panel} we give comparative example results for a range of dominant and sub-dominant modes for each boson spin, as a function of the dimensionless coupling $\alpha = \mu M_{\rm BH}$, with a fixed BH spin value, $a_* = 0.5$. In the \emph{right panel} we show example results for each spin value for the instability rates as a function of the dimensionless spin $a_*$ for a fixed value of the dimensionless coupling, $\alpha = 0.1$.}
  \label{fig:ratesplots}
\end{figure*}
To leading order the eigenfreqencies of the modes of the \emph{gravitational atom} are\footnote{See Ref.~\cite{Baumann:2018vus} for examples of higher-order corrections to the eigenfrequencies.},  
\begin{equation}
    \omega_{\rm R} \approx \mu \left[1-\frac{1}{2}\left(\frac{\alpha}{n+l+1}\right)^2\right] \sim \mu\ .
    \label{eq:realenergies}
\end{equation}

As the cloud is bosonic and not fermionic\footnote{Properties of degeneracy pressure with fermionic states ensure a large number of species are required to observe the superradiance process with fermionic species. See for example Ref.~\cite{Davoudiasl:2020uig} for superradiance bounds on this number.} in nature, the hydrogenic analogy of Eq.~(\ref{eq:realenergies}) contains an important addition from ingoing boundary conditions for waves infalling at the BH horizon. The true occupation numbers of the energy eigenstates are determined by quasinormal states/complex eigenfrequencies of the form, 
\begin{equation}
    \omega = \omega_{\rm R} + i \Gamma_{\rm SR} \ .
    \label{eq:supstates}
\end{equation}
The value of the small imaginary component, $\Gamma_{\rm SR}$, determines the nature of the bosonic clouds growth ($\Gamma_{\rm SR}>0$) or decay ($\Gamma_{\rm SR}<0$). The magnitude of this value gives the temporal change in amplitude from the dissipative boundary conditions at the horizon which occurs over the superradiance e-folding timescale, $\tau_{\rm SR} = \sfrac{1}{|\Gamma_{\rm SR}|}$. The resultant state upon saturation of Eq.~(\ref{eq:condition}) is a long lived bosonic cloud surrounding the BH spacetime. 

In the free-field limit the occupancy of each level of the system grows according to the superradiance rate until a quasi-stationary state is formed with the boson cloud, 
\begin{equation}
    \frac{dN}{dt}\Bigr|_{\rm SR} = \Gamma_{\rm SR}N\ ,
\end{equation}
which reaches maximum occupancy at \cite{Arvanitaki:2010sy,PhysRevD.91.084011},
\begin{equation}
    N_{\rm Max} \simeq \frac{G_{\rm N} M^2_{\rm BH}}{m}\Delta a_* \approx 10^{76} \times \left(\frac{\Delta a_*}{0.1}\right)\left(\frac{M_{\rm BH}}{10M_{\odot}}\right)^2 \ .
    \label{eq:nmax}
\end{equation}
For constraints on bosonic masses it is required the bosonic field should deplete the spin of the BH ($\Delta a_* \sim \mathcal{O}(1)$) within the relevant characteristic timescale of the BH, $\tau_{\rm BH}$,
\begin{equation}
    \Gamma_{\rm SR}\tau_{\rm BH}\geq \ln N_{\rm Max}\ .
    \label{eq:rateeq}
\end{equation}
  To explore the phenomenology of bosonic fields a common conservative instability timescale is the Salpeter time, $\tau_{\rm Sal}\simeq 4.5\times10^{7}$ Yrs \cite{1964ApJ...140..796S}, motivated by accretion models, where the compact object is radiating at it Eddington limit. A further conservative extension to this accounts for periods of super-Eddington accretion \cite{Baryakhtar:2017ngi} which defines $\tau_{\rm SEdd} = \sfrac{\tau_{\rm Sal}}{10} \simeq 4.5\times 10^{6}$ Yrs. Alternatively it may also be possible the Eddington accretion rate, $\dot{M}_{\rm Edd}$ is suppressed sufficiently such that the relevant BH evolutionary timescale is much longer (see Section~\ref{sec:dmbounds}). To allow for this we also consider instability timescales shorter than the Hubble time, $\tau_{\rm Hub} \sim 10^{10}$ Yrs. 

The superradiance process may offer a number of interesting signatures indicating the presence of ultralight bosons when the mass of the bosonic cloud is sufficient enough backreaction effects must be accounted for. This could lead to contributions to the stocstic background of GW detectable by LIGO/Virgo, emit monochromic GWs or be detectable from non-linear features in the evolution (see Section~\ref{sec:bosenova}), such as GW burts  triggered from self-interactions \cite{Arvanitaki:2010sy,Arvanitaki:2016qwi,PhysRevD.91.084011,Siemonsen:2019ebd,PhysRevD.96.064050}. 

A further possibility and the focus of this work is the exclusion of bosonic masses through notable BH absences in the BH mass-spin plane or \emph{Regge} plane. Each eigenmode is used to determine an exclusion isocontour window, where any BH spin measurement with sufficient spin which transcends this bound, constitutes an exclusion for a fixed boson mass. This makes the assumption the BH has failed to spin down over a sufficient time period due to the presence of an instability. Therefore a statistical investigation of whether current BH observations sit within these regions can lead to large exclusion windows over the ultralight domain. In Fig.~\ref{fig:regges} we present example Regge exclusion planes for both stellar mass BHs and SMBHs at fixed boson masses, many of which currently show strong evidence for exclusions from the presented BH data points. From this point forward we will normalise dimensionful constants and work in the units $G_{\rm N} = c = \hbar = 1$ and the value $M_{\rm Pl} = 2.435 \times 10^{18}\ {\rm GeV}$ represents the reduced Planck mass. We now review previous analytical results in the literature for the instability rates for each bosonic field spin we use to calculate exclusions. For an extensive review of superradiance see Ref.~\cite{Brito:2015oca}.

\subsection{The Free-Field Domain}
\label{sec:freefielddomain}

In the limit where the self-interaction strength between the massive 
bosons is sufficiently weak, the instability occurs at a rate according to Eq.~(\ref{eq:rateeq}), until Eq.~(\ref{eq:condition}) ceases to hold. Analytical solutions for $\Gamma_{\rm SR}$ are known in each case for linearised spin-0, -1 and -2 fields evolving on a fixed background, allowing bounds to be placed on their masses from astrophysical observations of BHs. Results are found from an understanding of the wave-functions behaviour, where approximate solutions are considered in different regimes of validity, in regards to the BH horizon. These methods lead to functions which suitably emulate the true numerical solutions at all radii. Below we summarise the results in the literature for each boson spin value we consider. See Ref.~\cite{Brito:2015oca} for a unified description of analytical results for superradiant instabilities in the linearised regime.

\subsubsection{Bosonic Fields of Spin-0}
\label{sec:spin0fields}

Massive scalar fields on a spacetime $g_{\mu\nu}$, obey the Klein-Gordon equation of motion,
\begin{equation}
    (g^{\mu\nu}\nabla_{\mu}\nabla_{\nu}- \mu_{0}^2 )\Phi=0\ .
\end{equation}
A vital property of the Klein-Gordon equation is its separability in Boyer-Lindquist coordinates via a general field ansatz of the form, $\Phi(t,\mathbf{r})=e^{-i\omega t+im\phi} R(r)S(\theta)$. Under the Teukolsky formalism \cite{Teukolsky:1973ha,PhysRevLett.29.1114} the Klein Gorden wave-equation separates into a function of solvable spheroidal wavefunctions. The relevant radial equation is a confluent Heun equation leading to a series of quasi-bound state solutions under the conditions of purely ingoing waves at the event horizon ($r\rightarrow r_+$) that vanish at spacial infinity ($r\rightarrow \infty$). Under a perturbative treatment of $\alpha$, asymptotic solutions are matched in an overlap region leading to analytical results for the frequency eigenvalues, $\omega$ in Eq.~(\ref{eq:supstates}). The validity of these solutions in the low coupling limit has been confirmed sufficient from exact numerical solutions (generally of the order $\alpha \lesssim \mathcal{O}(0.1)$) \cite{Cardoso:2005vk,Dolan:2007mj}. Analytical results are also available in the strong coupling limit ($\alpha \gg 1$) for scalars through a WKB analysis \cite{Zouros:1979iw}.
\begin{figure*}[!tbp]
  \centering
  \begin{minipage}[b]{0.49\textwidth}
    \includegraphics[width=\textwidth]{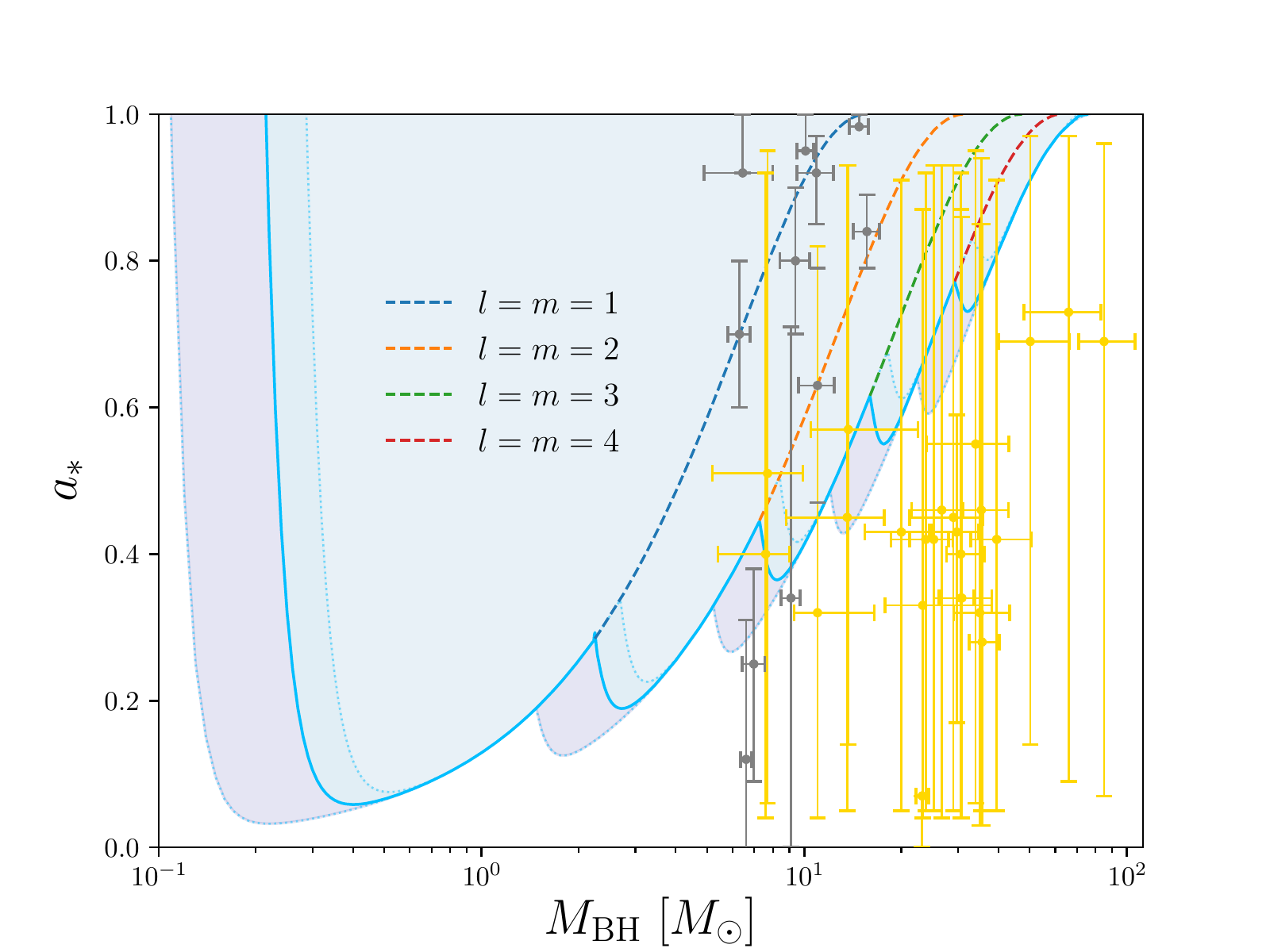}
  \end{minipage}
  \hfill
  \begin{minipage}[b]{0.49\textwidth}
    \includegraphics[width=\textwidth]{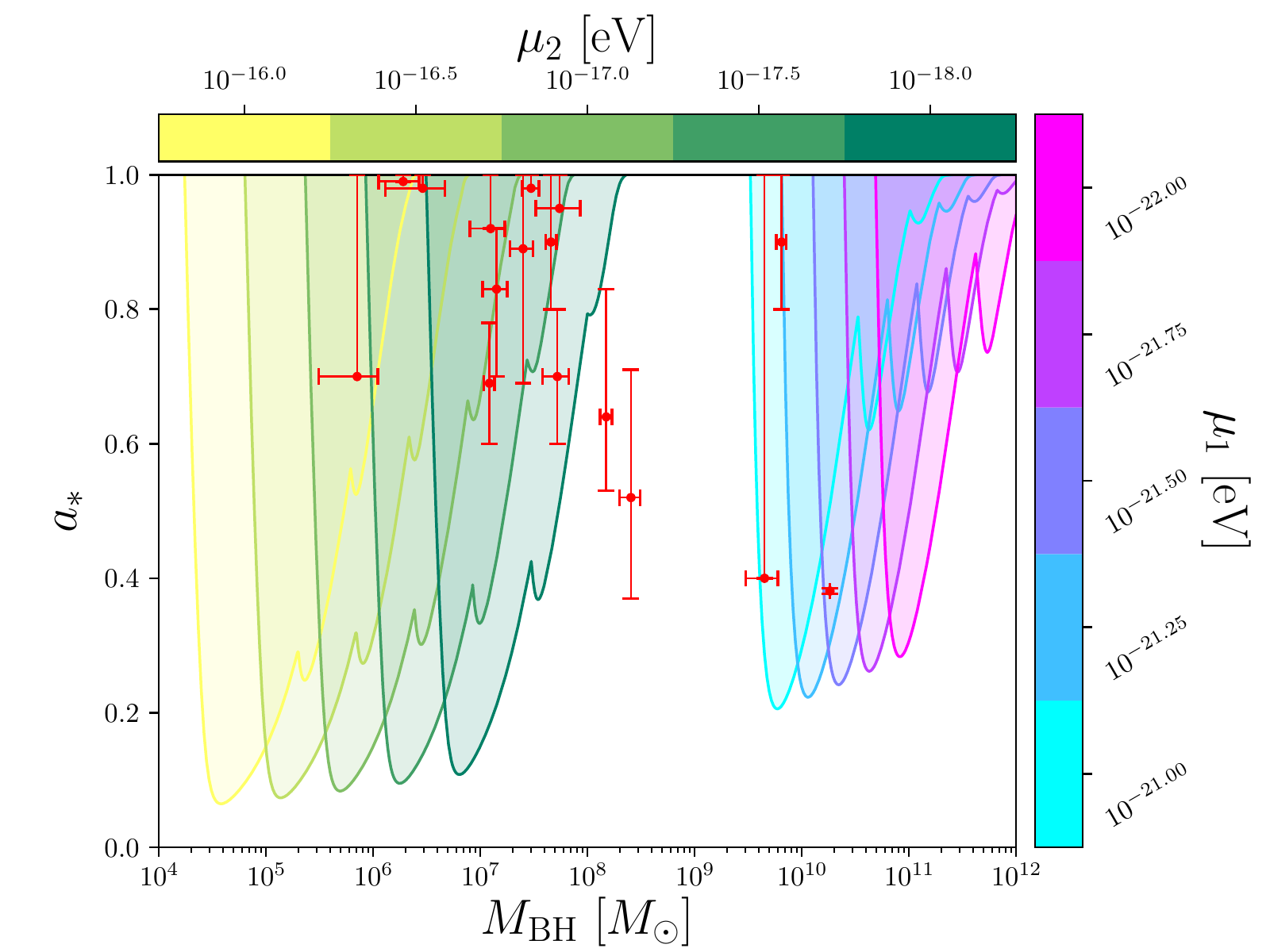}
  \end{minipage}
  \caption{Isocontour exclusion bounds in the BH mass-spin Regge plane for fixed bosonic masses. In the \emph{left panel} we show the stellar mass BHs with exclusion bounds from a spin-0 field of fixed mass, $\mu_0 = 4.3\times 10^{-12}\ {\rm eV}$, using the dominant modes with orbital/azimuthal quantum numbers equal to $l = m = (1\text{-}5)$. The grey/gold data points represent X-ray binary systems/ GW binary mergers from Table~\ref{tab:stellarBHs}/Table~\ref{tab:ligodata}. The \emph{darker outer}/\emph{lighter inner} shaded bounds come from using an instability timescale of the order of a Hubble time, $\tau_{\rm SR} = 1 \times 10^{10}$ Yrs and accounting for super-Eddington accretion, $\tau_{\rm SR} = 4.5 \times 10^{6}$ Yrs. These timescales also coincide with the longest and shortest timescales determined for stellar-mass binary merger scenarios \cite{Dominik:2013tma,2015ApJ...800....9M,TheLIGOScientific:2016htt}. In the \emph{right panel} we show the SMBH domain, giving example exclusion regions for fixed spin-1 and spin-2 masses using the Salpeter instability time scale, $\tau_{\rm SR} = 4.5 \times 10^7$ Yrs. The \emph{red} points represents the SMBHs found in Table~\ref{tab:SMBH}.  }
  \label{fig:regges}
\end{figure*}
For scalar states the analytical superradiance timescale as a solution to an eigenvalue problem is determined by \emph{Detweiler’s approximation} \cite{PhysRevD.22.2323}, 
\begin{align}
\Gamma^{nlm}_{0,\rm SR} &= 2\mu_{0}r_+(m\Omega_{\rm H}-\omega_{nlm})\alpha^{4l+4}\mathcal{A}_{nl}\mathcal{X}_{lm}\ , \label{eq:spin0rate} \\
\mathcal{A}_{nl} &= \frac{2^{4l+2}(2l+n+1)!}{(l+n+1)^{2l+4}n!}\left(\frac{l!}{(2l)!(2l+1)!}\right)^2\ , \\ 
\mathcal{X}_{lm} &= \prod^{l}_{k=1}(k^2(1-a_*^2)+4r_+^2(m\omega_{nlm}-\mu_0)^2\ .
\end{align}
The fastest growing level occurs for the nodesless ($n=2$) mode with the lowest orbital quantum number, which is the dipole mode, $\ket{nlm}=\ket{211}$. To leading order the growth rate of this mode is, $\Gamma^{211}_{\rm SR}\sim 24^{-1}a_*\alpha^8\mu$. Modes with $l\neq m$ are heavily suppressed compared to those with $l=m$ as shown in the \emph{left panel} of Fig.~\ref{fig:ratesplots} with the example mode $\ket{321}$.

\subsubsection{Bosonic Fields of Spin-1}
\label{sec:spin1fields}

The introduction of non-vanishing spin to a perturbing massive field adds technical complications in understanding its nature on the BH background. Massive vector fields on a spacetime $g_{\mu\nu}$, obey the Proca equations of motion, 
\begin{equation}
    \nabla_{\mu}F^{\mu\nu} = \mu_1^2A^{\nu}\ ,
\end{equation}
with Proca field strength, $F_{\mu\nu} = \partial_{\mu}A_{\nu}-\partial_{\nu}A_{\nu}$ and vector potential, $A_{\mu}$. The massive vector field explicitly breaks $U(1)$ symmetry, disregarding any gauge freedom in the vector potential and therefore satisfying the required Lorenz gauge conditions, $\nabla^{\mu}A_{\mu}$. The field equations then reduce to a description of four Klein-Gordon equations (i.e. $(\Box -\mu^2_1)A^{\mu}=0$). The Proca equation does not present the same \emph{neat} separability qualities (i.e. susceptible to the standard Teukolsky formalism as with the spin-0 case), whereby only recently did novel methods (FKKS ansatz \cite{Frolov:2018pys}) allow for a complete separation result, leading to a system of five solvable second-order PDEs. These results have subsequently been extended to the full Kerr-NUT-(A)dS family of BH spacetimes \cite{Krtous:2018bvk,Dolan:2018dqv} as well as Kerr-Newmen/Kerr-Sen solutions \cite{Cayuso:2019ieu}. Previous results for a minimally coupled massive vector fields used to define the linearised mode equations consist of semi-analytical methods conducted in the slow-rotation limit \cite{Pani:2012vp}, numerical time evolution solutions \cite{Witek:2012tr,East:2017ovw,East:2018glu} and numerical frequency-domain solutions \cite{Cardoso:2018tly}. 

Analytical results have been formulated using the method of matched-asymptotics in the non-relativistic limit ($\alpha \ll 1$), reproducing a set of solvable Schr\"{o}dinger type equations  \cite{Endlich:2016jgc,Baryakhtar:2017ngi,Baumann:2019eav}. The boundstate rate solutions for spin-1 fields are given as, 
\begin{align}
\Gamma^{nljm}_{1,\rm SR} &= 2\mu_{1}r_+(m\Omega_{\rm H}-\omega_{nlm})\alpha^{2l+2j+5}\mathcal{B}_{nlj}\mathcal{Y}_{jm}\ ,\label{eq:vectorrate} \\
\mathcal{B}_{nlj} &= \frac{2^{2l+2j+1}(l+n)!}{n^{2l+4}(n-l-1)!}\left(\frac{l!}{(l+j)!(l+j+1)!}\right)^2 \\
&\times \left(1+\frac{2(1+l-j)(1-l+j)}{(l+j)}\right)^2\ , \\
\mathcal{Y}_{jm} &= \prod^{j}_{k=1}(k^2(1-a_*^2)+4r_+^2(m\omega_{nlm}-\mu_0)^2\ ,
\end{align}
valid for the mode set, $\{j=l+1;l\}$. The dominant mode for vectors is $\ket{nljm}=\ket{1011}$, with the approximate scaling proportionality, $\Gamma_{\rm SR}^{1011}\propto 4a_*\alpha^6\mu_{1}$ \cite{Rosa:2011my,Pani:2012bp,Pani:2012vp,East:2017mrj,East:2017ovw,Baryakhtar:2017ngi,Cardoso_2018,Dolan:2018dqv,Siemonsen:2019ebd,Baumann:2019eav} as shown in the \emph{left panel} of Fig.~\ref{fig:ratesplots}. Example solutions to Eq.~(\ref{eq:vectorrate}) as a function of the BH spin are shown in the \emph{right panel} of Fig~\ref{fig:ratesplots}. 

\subsubsection{Bosonic Fields of Spin-2}
\label{sec:spin2fields}
The separability of massive tensor field perturbations on the Kerr background has only recently been explored in the context of superradiant instabilities. Considering a general curved spacetime, it has been shown at the linear level there exists consistent field equations for massive spin-2 fields \cite{Brito:2013wya,Mazuet:2018ysa} which lead to the emergence of a description of superradiant instabilities for unstable hydrogenic states \cite{Brito:2013yxa,Brito:2013wya,Brito:2015oca,PhysRevLett.124.211101}. Previous results in the literature have produced considerations in the slow-rotation approximation to first order in the Kerr parameter \cite{Brito:2013wya} and solutions in the limit, $\alpha \ll 1$, for any values of the dimensionless spin \cite{PhysRevLett.124.211101}. At leading order the linearised field equations for a massive spin-2 test field are defined as \cite{Brito:2013wya},  
\begin{align}
   \Box H_{\alpha\beta} + 2R_{\alpha\gamma\beta\delta}H^{\gamma\delta} + \mu_2^2H_{\alpha\beta} &= 0\ , \\
   \nabla^{\alpha}H_{\alpha\beta} = 0, \quad H^{\alpha}_{\alpha}&=0 \ ,
\end{align}
where $H_{\alpha\beta}$ represents a canonically normalised mass eigenstate for a spin-2 field and $R_{\alpha\gamma\beta\delta}$ is the Riemann tensor.
Using Eq.~(A25) in Ref.~\cite{PhysRevLett.124.211101} the authors analytically compute the instability time scales for dominant modes by utilising the BH absorption probability for long-wavelength massless spin-2 waves. To leading order the absorption probability and the Schwarzschild decay rate are used to determine the spin dependence on the decay rate for the relevant quasi-bound state \cite{Baryakhtar:2017ngi}. The superradiance rates in the spin-2 case are defined as, 
\begin{align}
\Gamma_{2,\rm SR}^{nljm} &=  \mathcal{T}_{lj} (m\Omega_{\rm H} - \omega_{nljm})\alpha^{2l+2j+5} \left(\frac{\mathscr{C}_{jm}(a_*)}{\mathscr{C}_{jm}(0)}\right)\ , \label{eq:spin2rate} \\
\mathscr{C}_{jm}(a_*) &= (1+\Delta)\Delta^{2j}\mathcal{Z}_{nljm} \ , \\
\mathcal{Z}_{nljm} &= \prod^{j}_{k=1}\left[1+4M_{\rm BH}^2\left(\frac{(\omega_{nljm}-m\Omega_{\rm H})}{k\Delta(1+\Delta)^{-1}}\right)\right] \ ,
\end{align}
where $\Delta = \sqrt{(1-a_*^2})$ is the BH absorption probability \cite{PhysRevLett.124.211101}. These solutions are valid for the mode set, $\{ j\in(l-2,l+2)\geq 0, m\in(-j,j)\}$ and the numerical constants $ \mathcal{T}_{lj}$ can be found in Table~I of Ref.~\cite{PhysRevLett.124.211101}. Unlike the case of spin-0 and spin-1 fields, spin-2 fields possess two dominant states, $\ket{n,j,l,m}$, through the non-axisymmetric mode requirements, $m\neq 0$ so $j\geq 1$. These are the dipole mode, $j=l=1$ and quadrupole mode, $j=2,l=0$. As pointed out in Refs.~\cite{Brito:2015oca,PhysRevLett.124.211101} the existence of a special, potentially dominant dipole mode \cite{Brito:2013wya} is distinct in the small coupling limit whilst not possessing an overtone. Current solutions for this mode are only provided to first order in the spin. We therefore follow the the work of Refs.~\cite{PhysRevLett.124.211101,Brito:2015oca} and only consider the subsequent leading order modes for our constraints. This means the results in Section~\ref{sec:frefieldresults} might be conservative in terms of not accounting for instability time-scales from modes shorter than the case of spin-1 fields etc. \cite{Brito:2013wya}.

\subsection{Nonlinear-self-interactions and Bosenova}
\label{sec:bosenova}
There are two principle specific non-linear phenomena which could significantly effect the exponential growth of the bosonic cloud. These are level mixing between superradiance levels and the presence of a series of dynamical collapses known as a \emph{bosenova} \cite{Arvanitaki:2010sy,PhysRevD.91.084011,Yoshino:2012kn,Yoshino:2015nsa,Fukuda:2019ewf} from self-interactions within the cloud. For isolated systems level mixing effects may also quench the evolution of the instability as the BH could sits on a Regge trajectory during the spin down process. We comment on this in Appendix~\ref{app:signals}. 
For binary systems the presence of a secondary component (NS or BH etc.) may also have a significant effect on the profile of the boson cloud, and the rate at which the instability evolves through inspiral dynamics \cite{Baumann:2018vus,Zhang:2019eid,PhysRevD.99.064018,Baumann:2019ztm,Berti:2019wnn,Wong:2020qom,Kavic:2019cgk,Cardoso:2020hca}. Tidal effects such as resonances can lead to significant deformations of the hydrogenic wavefunctions via transitions between growing and decaying eigenmodes \cite{Baumann:2018vus,Berti:2019wnn}. This may also have significant effects for observational GW astronomy \cite{Hannuksela:2018izj,Baumann:2019ztm} (sinking orbits etc. \cite{Cardoso:2011xi,Zhang:2018kib,Zhang:2019eid}). 

For the analysis in this work we adopt the simplification of considering the evolution of the cloud to be dominated by a singular eigenmode with the other eigenmodes evolving independently on the Kerr background, representing a decoupled series of quasi-stationary states. In the case of merger measurements (Table~\ref{tab:ligodata}) we assume that the instability is effective throughout the binary systems lifetime. This would relate to the approximation that the BHs are well separated before the merger and the approximation of the instability being allowed to evolve independently before the merger event occurs. See Ref.~\cite{Fernandez:2019qbj} for an analysis with both cases.  

\begin{figure*}[!tbp]
  \centering
  \begin{minipage}[b]{0.49\textwidth}
    \includegraphics[width=\textwidth]{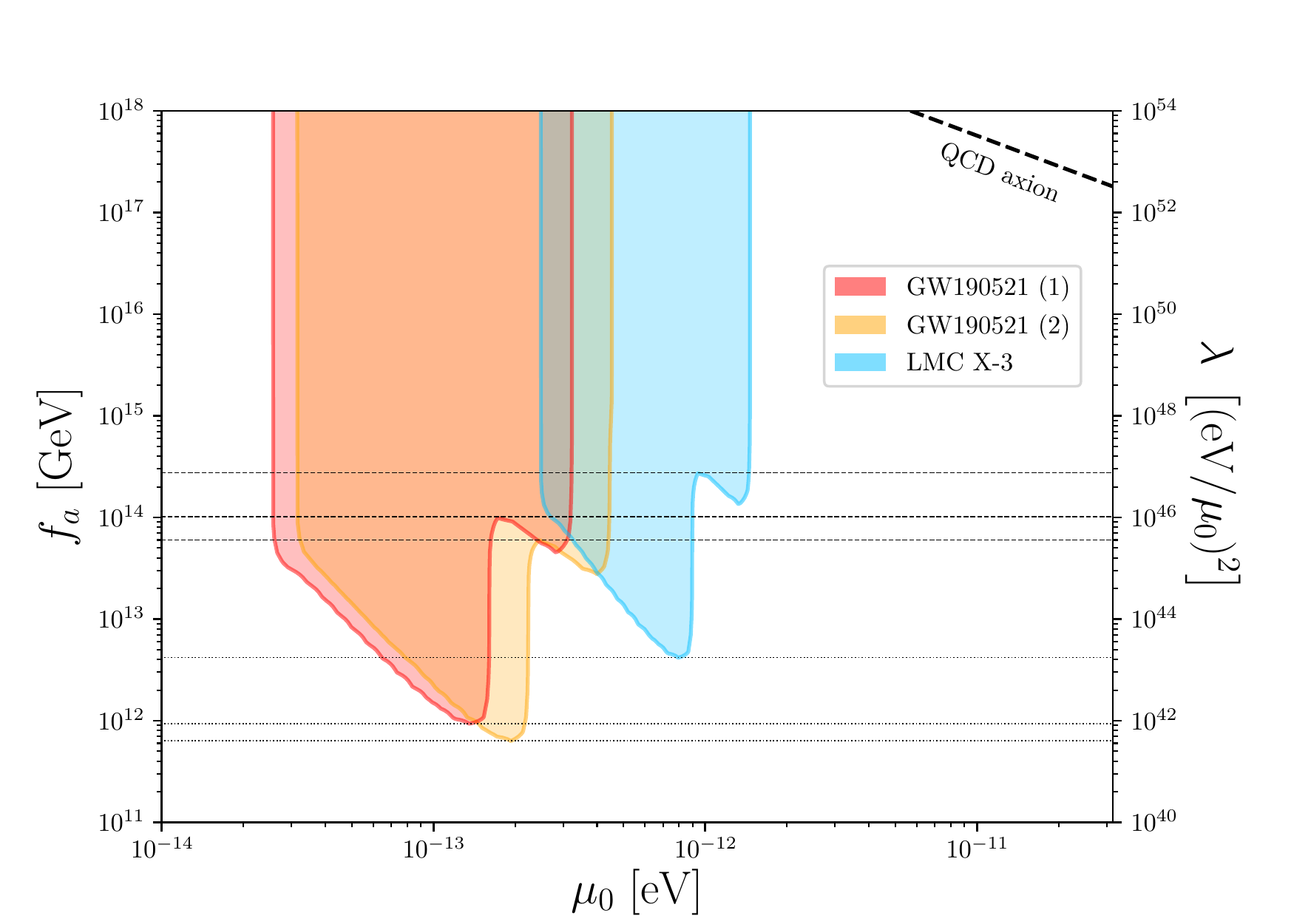}
  \end{minipage}
  \hfill
  \begin{minipage}[b]{0.49\textwidth}
    \includegraphics[width=\textwidth]{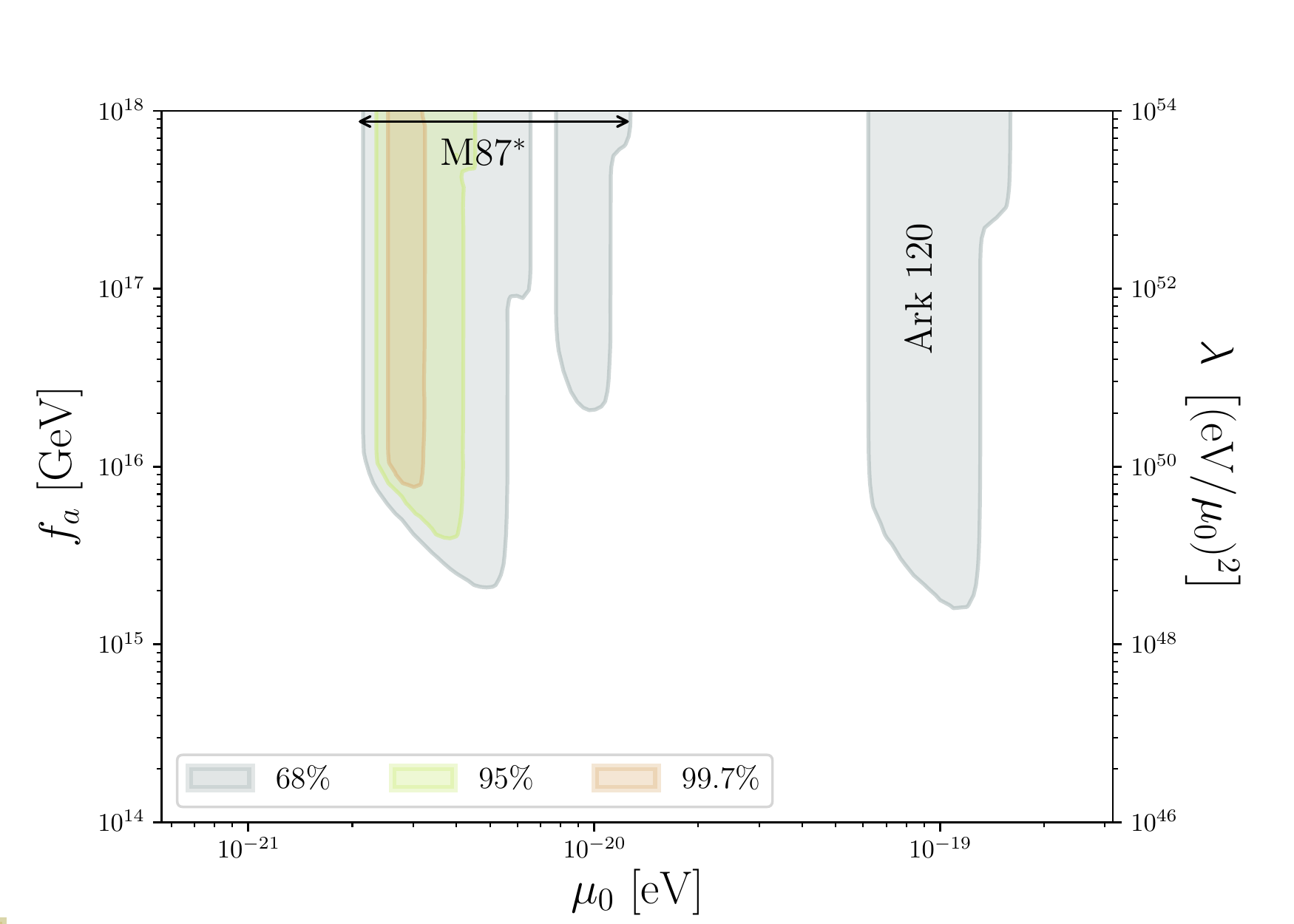}
  \end{minipage}
  \caption{Constraints on the axion decay constant/spin-0 field self-coupling strength and masses determined from instability timescales calculated using Eq.~(\ref{eq:boserate}). In each example the bounds drop out due to the occurrence of a bosenova before the instability extracts maximal spin from the BH. In the \emph{left panel} we show example bounds at the 68\% confidence interval for stellar mass BHs, specifically the primary and secondary components of GW190521 and X-ray binary, LMC X-3. In the \emph{right panel} we display bounds on the lightest spin-0 fields excluded from the some of the heaviest BHs in our ensemble. In particular M87* is able to constrain masses approaching the limits for FDM up to $f_a\sim 10^{15.5}$ GeV. }
  \label{fig:boseplots}
\end{figure*}

Couplings to standard matter components can also suppress the superradiant growth, such as couplings to photons \cite{PhysRevLett.122.081101,Boskovic:2018lkj}. Alternatively for a general (pseudo-)scalar boson, $\phi$, described by a periodic potential of the form, $U=f^2_a\mu^2_0(1-\cos(\sfrac{\phi}{f_a}))$, the self-interaction strength ($\lambda \equiv \sfrac{\mu^2_0}{f^2_a}$) of the field is governed by the scale of the spontaneously broken U(1) symmetry of the theory. The nature of the quartic coupling ($\lambda \sfrac{\phi^4}{4!}$) can be prohibitive to the superradiant evolution when the fields decay constant is of the order $f_a\ll M_{\rm GUT} \approx 10^{16}$ GeV \cite{PhysRevD.91.084011}. In this case the hydrogenic wave functions are no longer valid when the self-interactions within the Bose–Einstein condensate overcome the gravitational binding energy ($\Phi\sim f_a$), leading to a rapid collapse of the cloud. The critical occupation number at which this occurs is \cite{PhysRevD.91.084011}, 
\begin{equation}
    N_{\rm Bose} \simeq c_0 10^{78}\left( \frac{n^4}{\alpha^3}\right)\left(\frac{M_{\rm BH}}{10M_{\odot}} \right)^2 \left(\frac{f_a}{M_{\rm Pl}} \right)^2\ .
    \label{eq:nbose}
\end{equation}
The numerical constant $c_0 \sim 5$ is determined through numerical simulations \cite{Yoshino:2012kn}. The interplay between the values of Eq.~(\ref{eq:nbose}) and Eq.~(\ref{eq:nmax}) determine the efficiency of the superradiance process, where the characteristic timescale in which the instability extracts sufficient spin within a single cycle, before self-interactions become significant comes from a now modified version of Eq.~(\ref{eq:rateeq}),
\begin{equation}
\Gamma_{\rm SR} \tau_{\rm BH} (\sfrac{N_{\rm Bose}}{N_{\rm Max}}) > \ln N_{\rm Bose}\ .
\label{eq:boserate}
\end{equation}
Using Eq.~(\ref{eq:boserate}) we can place exclusion bounds in the self-interaction/mass plane for bosons. We present these results in Section~\ref{sec:interactingresults} and Appendix~\ref{app:extendedinteracting}.

\section{Bosonic Bounds from Black Hole Spin Measurements}
\label{sec:bounds}

Generally measurements of BH spins are often susceptible to large uncertainties. We will account for this by giving both results for bounds from individual BHs and the total ensemble formed from the data in Appendix~\ref{app:bhdata}. This ensemble represents a population of astrophysical BHs inside the mass range, $5M_{\odot} \lesssim M_{\rm BH} \lesssim 10^{10}M_{\odot}$. We adopt the methodology of Ref.~(\cite{Stott:2018opm}) (see Appendix~B) for our calculations. We follow this by stating bounds from the total ensemble on the self-coupling strength of ultralight scalars. 

\subsection{The Free-Field Domain}
\label{sec:frefieldresults}
The following results apply to bosons where we consider only their gravitational interactions. In Table~\ref{tab:68percentresults} we present the total exclusion window for each spin at the 68\% confidence limit. Likewise in Table~\ref{tab:95percentresults} we present the total exclusion window for each spin at the 95\% confidence limit. The following bounds are calculated using the instability rates in Eq.~(\ref{eq:spin0rate}), Eq.~(\ref{eq:vectorrate}) and Eq.~(\ref{eq:spin2rate}) for spin-0, spin-1 and spin-2 fields respectively. We use the total BH data ensemble formed of Table~\ref{tab:stellarBHs}, Table~\ref{tab:ligodata} and Table~\ref{tab:SMBH} with a BH timescale equal to the Salpeter time. For completeness the individual bounds for each BH using the instability timescales, $\tau_{\rm SEdd}$, $\tau_{\rm Sal}$ and $\tau_{\rm Hub}$, detailed in Section~\ref{sec:superradiance}, can be found in Appendix~\ref{app:extendedfree}. These are presented in Table~\ref{tab:extendedxray} (X-ray binaries), Table~\ref{tab:extendedligo} (GW binary mergers) and Table~\ref{tab:extendedsuper} (SMBHs).

\begin{table}[t]
  \centering
    \caption{Exclusion windows for massive bosonic fields with integer spin, 0, 1 and 2 determined using the analytical rates defined in Section~\ref{sec:freefielddomain} at the 68\% confidence level. }
  \resizebox{0.48\textwidth}{!}{%
  \begin{tabular}{@{}c|c@{}}
    \toprule
    Boson Spin  & 68\% Confidence Limit Mass Bounds     \\
    \midrule
   Spin-0 &\begin{tabular}{@{}c@{}c@{}}  \footnotesize $3.8\times 10^{-14}\ \text{eV}   \leq   \mu_{0} \leq 3.4\times 10^{-11}\ \text{eV}  $ \\\footnotesize$5.5\times 10^{-20}\ \text{eV} \leq    \mu_{0} \leq 1.3\times 10^{-16}\ \text{eV} $ \\\footnotesize$2.5\times 10^{-21}\ \text{eV} \leq    \mu_{0} \leq 1.2\times 10^{-20}\ \text{eV}$\end{tabular} \\ \midrule \midrule Spin-1 & \begin{tabular}{@{}c@{}c@{}}\footnotesize$6.2\times 10^{-15}\ \text{eV}   \leq   \mu_{1} \leq 3.9\times 10^{-11}\ \text{eV}$ \\\footnotesize $2.8\times 10^{-22}\ \text{eV} \leq    \mu_{1} \leq 1.9\times 10^{-16}\ \text{eV} $\end{tabular} \\ \midrule \midrule Spin-2 & \begin{tabular}{@{}c@{}c@{}}\footnotesize$2.2\times 10^{-14}\ \text{eV}   \leq   \mu_{2} \leq 2.8\times 10^{-11}\ \text{eV}$ \\\footnotesize$1.8\times 10^{-20}\ \text{eV}  \leq    \mu_{2} \leq 1.8\times 10^{-16}\ \text{eV}$ \\\footnotesize$6.4\times 10^{-22}\ \text{eV} \leq    \mu_{2} \leq 7.7\times 10^{-21}\ \text{eV} $\end{tabular} \\ 
        \midrule
    \bottomrule
  \end{tabular}}
      \label{tab:68percentresults}
\end{table}

\begin{table}[t]
  \centering
    \caption{Exclusion windows for massive bosonic fields with integer spin, 0, 1 and 2 determined using the analytical rates defined in Section~\ref{sec:freefielddomain} at the 95\% confidence level. }
  \resizebox{0.48\textwidth}{!}{%
  \begin{tabular}{@{}c|c@{}}
    \toprule
    Boson Spin  & 95\% Confidence Limit Mass Bounds    \\
    \midrule
   Spin-0 &\begin{tabular}{@{}c@{}c@{}}  \footnotesize $4.3\times 10^{-14}\ \text{eV}   \leq   \mu_{0} \leq 2.7\times 10^{-11}\ \text{eV} $ \\\footnotesize$1.7\times 10^{-19}\ \text{eV}   \leq   \mu_{0} \leq 5.9\times 10^{-17}\ \text{eV} $ \\\footnotesize$2.7\times 10^{-21}\ \text{eV}   \leq   \mu_{0} \leq 4.5\times 10^{-21}\ \text{eV} $\end{tabular} \\ \midrule \midrule Spin-1 & \begin{tabular}{@{}c@{}c@{}}\footnotesize$6.5\times 10^{-15}\ \text{eV}   \leq   \mu_{1} \leq 2.9\times 10^{-11}\ \text{eV} $ \\\footnotesize$2.9\times 10^{-22}\ \text{eV} \leq    \mu_{1} \leq 1.2\times 10^{-16}\ \text{eV} $\end{tabular} \\ \midrule \midrule Spin-2 & \begin{tabular}{@{}c@{}c@{}}\footnotesize$2.5\times 10^{-14}\ \text{eV}   \leq   \mu_{2} \leq 2.2\times 10^{-11}\ \text{eV}$ \\\footnotesize$3.1\times 10^{-20}\ \text{eV}   \leq   \mu_{2} \leq 9.1\times 10^{-17}\ \text{eV}$ \\\footnotesize$6.4\times 10^{-22}\ \text{eV}   \leq   \mu_{2} \leq 7.7\times 10^{-21}\ \text{eV} $\end{tabular} \\ 
        \midrule
    \bottomrule
  \end{tabular}}
      \label{tab:95percentresults}
\end{table}

\subsection{The Interacting Domain}
\label{sec:interactingresults}

In the interacting domain the self-interaction of the boson condensate is sufficient to interrupt the superradiance process, which generally occurs at the order $f_a \leq M_{\rm GUT} \sim 10^{16}\ {\rm GeV}$. In Fig.~\ref{fig:boseplots} we display example bounds on the self-interaction and masses of ultralight scalars. In the \emph{left panel} we show bounds at the 68\% confidence interval from the two BHs detected in the LIGO event GW190521 as well as X-ray binary LMC X-3. As $f_a \rightarrow M_{\rm Pl}$ the constraints approach the free-field results (see Table~\ref{tab:extendedxray} and Table~\ref{tab:extendedligo}). These bounds represent constrains on the self-interactions of the lightest bosons constrained from stellar mass BHs. The \emph{black dashed} lines ($f_a \sim 10^{14}\ {\rm GeV}$) represent the lowest value for $f_a$ where the exclusion bounds in the free-field limit for the mass cease to hold from the shape of the isocontours in the Regge plane, determined by higher order modes. The \emph{black dotted} lines ($f_a \sim 10^{12}\ {\rm GeV}$) represent the lowest value of $f_a$ constrained from superradiance which is peaked at the approximate value $\sim \bar{\mu}^{68\%}_0$, representing the mean value of the free-field bounds. In the \emph{right panel} we show example bounds from observations of M87* and Ark 120 (see Section~\ref{sec:dmbounds}) as constraints on models of ultralight axionic DM. The results for each BH detailed in Appexdix~\ref{app:bhdata} are given in Table~\ref{tab:extendeddecay}.

Combining the free-field results in Section~\ref{sec:frefieldresults} we can take the highest individual bound on $f_a$ as a conservative estimate of the value of $f_a$ bounded across the full mass exclusion Windows given in Table~\ref{tab:68percentresults}. To make this estimate we consider the value in which the inflection point of the first subdominant mode forms a orthogonal projection to the $f_a$ axis, as shown in the \emph{left panel} of Fig.~\ref{fig:boseplots}. If the first subdominant mode is close to $M_{\rm Pl}$ we quote the value in which first derivative of the outer contour bound is an approximately well defined quantity (where the minimum constrained mass is not a constant function of $f_a$) on the edge of the lowest constrained masses, in order to capture a sufficient area of the bounds. In the case of stellar mass X-ray binary sources (Table~\ref{tab:stellarBHs}) we can exclude the values,  
\begin{equation}
  f_{\rm a} \gtrsim 1.2 \times 10^{15}\ {\rm GeV} \ ,
\end{equation}
across the approximate mass interval, $9.9 \times 10^{-14} \ {\rm eV}\leq \mu_0 \leq 3.2 \times 10^{-11}\ {\rm eV}$. From GW merger data (Table~\ref{tab:ligodata}) the values, 
\begin{equation}
f_{\rm a} \gtrsim\ 2.5 \times 10^{14}\ {\rm GeV}\ ,
\end{equation}
over the approximate mass interval, $2.3 \times 10^{-14} \ {\rm eV}\leq \mu_0 \leq 2.1 \times 10^{-11}\ {\rm eV}$.
Likewise for the SMBH domain we find the bounds, 
\begin{equation}
f_{\rm a} \gtrsim\ 2.0 \times 10^{16}\ {\rm GeV}\ ,
\end{equation}
over the approximate mass interval, $2.5 \times 10^{-21} \ {\rm eV}\leq \mu_0 \leq 1.4 \times 10^{-16}\ {\rm eV}$. In regards to the bounds stated in Table~\ref{tab:68percentresults} for spin-0 bosons we can exclude $f_a \geq 1.5\times 10^{15} {\rm GeV}$ over the interval $3.8\times 10^{-14}\ {\rm eV}\leq \mu_0 \leq 3.4\times 10^{-11}\ {\rm eV}$, $f_a \geq 1.7\times 10^{16}  {\rm GeV}$ over the interval $5.5\times 10^{-20}\ {\rm eV}\leq \mu_0 \leq 1.3\times 10^{-16} \ {\rm eV}$ and $f_a \geq 2.0\times 10^{16}  {\rm GeV}$ over the interval $2.5\times 10^{-21}\ {\rm eV}\leq \mu_0 \leq 1.2 \times 10^{-20}\ {\rm eV}$, in concordance with previous results found in Refs.~\cite{Arvanitaki:2016qwi,PhysRevD.91.084011,Zu:2020whs}. 
 
\subsection{Black Hole Mass Gaps}
\label{sec:massgaps}
For constrains on ultralight bosons from BH mass-spin measurements there are four mass gaps which currently fix probable mass ranges. Quantifying the limits of these gaps will ultimately translate into defining hard edges for the upper and lower bosonic field constraints, and are therefore important in the context of phenomenological model viability. The first of these is the NS/BH or \emph{lower} mass gap \cite{Bailyn_1998,_zel_2010} which traditionally spans orders of $\sim[2\text{-}5] \ M_{\odot}$\footnote{
The current heaviest NS with robust measurements has a mass of $2.01\pm0.04 M_{\odot}$ \cite{Antoniadis:2013pzd}. See also PSR J0740+6620 ($2.14^{+0.10}_{0.09}M_{\odot}$) in Ref.~\cite{Cromartie:2019kug}} based on stellar evolution arguments and an absence of any well defined gravitational and/or electromagnetic wave measurements for either a BH or NS in this window \cite{2012ApJ...757...36K}. The second mass gap regarding stellar mass candidates is the \emph{high mass gap} from the physics of electron-positron pairs in stellar cores, known as pair-instability supernova (PISN) \cite{1964ApJS....9..201F,PhysRevLett.18.379,Woosley_2017} as well as pulsational pair-instability supernova (PPISN) \cite{Woosley:2016hmi} theory. This traditionally excludes masses defined by the lower bound $M_{\rm BH}\gtrsim (50\text{-}65) M_{\odot}$. The upper bound on the gap is defined for heavier objects which quench the pair-instability at the order, $M_{\rm BH}\sim 120-135 M_{\rm BH}$ through photo-disintegration. This limit also indicates the expected lower bound on IMBHs and the first ever direct observation of an IMBH appears to conform to this understanding but its merger component however do not \cite{Abbott:2020tfl}. Generally inference of the masses of current IMBH candidates is highly non-trivial (e.g see Refs.\cite{Lin:2020exl,Dong:2006mh,2015ApJ...809L..14B,2018ApJ...863....1C,2010ApJ...712L...1I}) but evidence of their presence is growing \cite{2019arXiv191109678G}, suggesting the BH mass function is a continuum from $\mathcal{O}(1)$ solar masses to billions of solar masses. It is expected IMBH measurments \cite{2019arXiv191109678G} ($M_{\rm BH}\sim (10^2\text{-}10^5)M_{\odot}$) will steadily fill this gap through next-generation experimentation, e.g. LISA \cite{Miller:2008fi,Barausse:2020rsu}.

Jumping the current IMBH gap, LMBHs represent the lightest objects associated to the general consensus many galaxies are expected to possess SMBHs \cite{Richstone:1998ky,2013ARA&A..51..511K}. An understanding of the histories of LMBHs could bring insight into the nature of seed-BHs, through indirect methods, considered as a suitable candidate to explain the formation of SMBHS \cite{2013ApJ...771..116J,Fan:2001ff}. Examples include the bulgeless galaxy NGC 4395 \cite{1989ApJ...342L..11F,Filippenko:2003kg} possessing an object with an infered mass of $M_{\rm BH}\sim 3 \times 10^{5}M_{\odot}$ \cite{2005ApJ...632..799P} or the dwarf spheroidal galaxy POX 52 \cite{2004ApJ...607...90B} ($M_{\rm BH}\sim 3\times 10^{5}M_{\odot}$\cite{2004ApJ...607...90B}). These BHs are often incorporated into the IMBH class where many candidates have been identified in local analyses, such as the Sloan Digital Sky Survey (SDSS) \cite{Greene:2005nj,2007ApJ...667..131G,2018ApJS..235...40L} or deeper searches \cite{2007ApJ...654..125S,2009ApJ...698.1515D,2011Natur.470...66R}. They can be difficult to study as their measurements are susceptible to large systematics from their apparent luminosity, but are key target for direct detection in future experiments through GW signatures etc.

Finally UMBHs represent the heaviest inferred observations to date, transcending masses of the order $\sim10^{10} M_\odot$. It has been argued there is a theoretical maximum redshift-independent value for the masses of UMBHs of the order $\sim 5\times 10^{10}M_{\odot}$ \cite{Natarajan:2008ks,Kormendy:2013dxa,2016MNRAS.456L.109K} (this extends to $3\times 10^{11}M_{\odot}$ for extremal prograde spin values). It could be argued BHs could be found well above this limit. So called \emph{Stupendously Large BHs} (SLABs) \cite{Carr:2020erq}, which my be either primordial in origin or sit in the high-mass tail of the UMBH population, could be as massive as $(10^{12}-10^{18})M_{\odot}$. Highly spinning objects ($a_*\gtrsim 0.9$) could lead to constraints on the masses of bosonic fields using the analytical solutions in Section~\ref{sec:freefielddomain}, of the order $\sim 10^{-25.5}\ {\rm eV}, \sim 10^{-27.5}\ {\rm eV}$ and $\sim 10^{-28}\ {\rm eV}$ for spin-0, -1 and -2 fields respectively. See Ref.~\cite{Carr:2020erq} for details and superradiance bounds from SLABs. Below we explore the possible bounds emerging from each of these gaps in the context of specific phenomenological models. The origins and life-cycles of these objects are of course heavily dependant on different classes of models. The following bounds from objects with uncertain properties are therefore only representative of possible limits conforming to the assumptions made in our analysis and are far from robust in nature.

\subsubsection{The Neutron Star-Black Hole desert and the QCD Axion}

There are several examples to date of unclassified objects which may fall in the \emph{lower mass gap}, capable of pushing the bounds on the heaviest bosonic fields constrained by superradiance. In particular this is of great interest to constraining the mass of the QCD axion \cite{PhysRevD.91.084011} whilst also ensuring $f_a^{\rm QCD} \lesssim M_{\rm Pl}$ from weak gravity conjecture \cite{ArkaniHamed:2006dz,Rudelius:2015xta,Bachlechner:2015qja,Montero:2015ofa}, entropy bounds \cite{Conlon:2012tz} or field range arguments \cite{Brown:2015lia} etc. Using the bounds for spin-0 fields in Table~\ref{tab:68percentresults} corresponds to the following bounds on the QCD axion with a sub-Planckian decay constant, $2.3\times 10^{-12} \lesssim \mu^{\rm QCD}_{0}\lesssim 3.4\times 10^{-11}\ {\rm eV}$ corresponding to $M_{\rm Pl}\gtrsim f^{\rm QCD}_{a}\gtrsim 1.7 \times 10^{17}\ {\rm GeV} $ and  $2.3\times 10^{-12} \lesssim \mu^{\rm QCD}_{0}\lesssim 2.7\times 10^{-11}\ {\rm eV}$ corresponding to $M_{\rm Pl}\gtrsim f^{\rm QCD}_{a}\gtrsim 2.1 \times 10^{17}\ {\rm GeV} $ for the 68\% and 95\% confidence limits respectively. 

It is generally expected NS masses should somewhat follow theoretical arguments for their maximum allowed values such as the Tolman–Oppenheimer–Volkoff (TOV) hydrostatic equilibrium limit \cite{1968ApJ...153..807H,Lattimer:2000nx}, for rigidly spinning objects, $M_{\rm TOV}\sim 2.5 M_{\odot}$ ($\sim2M_{\odot}$ in the non-rotating case). See example recent bounds on this value from  GW170817, $M_{\rm TOV} \sim 2.3M_{\odot}$ \cite{Shibata:2019ctb}, $M_{\rm TOV} \sim (2.16-2.28)M_{\odot}$ \cite{PhysRevD.97.021501} and $M_{\rm TOV} \sim 2.17M_{\odot}$ (\cite{Rezzolla:2017aly,Margalit:2017dij}). Assuming this as an approximate upper bound, then recently several interesting measurements have been made for compact objects difficult to classify without a detailed analysis of their structure and evolution, one example is the $M_{\rm BH}=3.3\substack{+2.8 \\ -0.7}M_{\odot}$ BH candidate residing in the binary system
2MASS J05215658+4359220 detailed in Ref.~\cite{Thompson637} (see also Refs.~\cite{vandenHeuvel:2020chh,Thompson:2020nbd} for discussions on this object).

Alternatively two rather mysterious system detections by LIGO/Virgo may hint at evidence of observed BHs sitting neatly above the mass of all known binary NS \cite{Ozel:2016oaf}, but significantly below any robust current stellar mass BH observations to-date. Firstly the binary system in the event GW190425 \cite{Abbott:2020uma} is unusual due to its observed mass ($3.4\pm^{0.3}_{0.1}M_{\odot}$) being significantly larger than previously known galactic double NS binaries \cite{Farrow:2019xnc,2019MNRAS.488.5020Z,Gupta:2019nwj}, and still it has not been fully ruled out if either one or both of these components could be a BH. Event GW190814 involves the most asymmetrical binary mass ratio to date, where its secondary object has a measured mass, $M_{\rm BH} = 2.59\pm^{0.08}_{0.09} M_{\odot}$. The absence of features such as an electromagnetic counterpart or measurable tidal deformation currently leave the nature of the secondary compact object also open for debate. Given the mass of the secondary object appears to follow $M_{\rm BH}> M_{\rm TOV}$ from current bounds it could be argued the system is indeed a binary BH merger event \cite{Carr:2020xqk,Jedamzik:2020omx,Clesse:2020ghq,Vattis:2020iuz}, however see Refs.~\cite{Tan:2020ics,Fattoyev:2020cws,Tsokaros:2020hli,Most:2020bba} for BH-NS arguments. Measuring the spin of the secondary object is challenging due to the asymmetry in the merger component masses. 

As an indication of what future observations in this area could present in terms of bounds for bosonic fields we use both the BH of Ref.~\cite{Thompson637} and the secondary component of GW190814 as examples. We adopt a proxy prior in each case of $a_* = 0.8\pm0.2$ representing a sufficiently high spin for susceptibility to superradiance and use the Salpeter instability timescale. Doing so gives the approximate bounds on ultralight bosonic fields  $4.0 \times 10^{-13} {\rm eV} \lesssim \mu_0 \lesssim 4.1 \times 10^{-11} {\rm eV} $, $5.9 \times 10^{-14} {\rm eV} \lesssim \mu_1 \lesssim 4.1 \times 10^{-11}  {\rm eV}$ and $2.3 \times 10^{-13}  {\rm eV} \lesssim \mu_2 \lesssim 3.4 \times 10^{-11}  {\rm eV}$ for spin-0, -1 and -2 respectively for the BH of Ref.~\cite{Thompson637}. For the secondary component of GW190814 we find $5.2 \times 10^{-13} {\rm eV} \lesssim \mu_0 \lesssim 5.3 \times 10^{-11} {\rm eV} $, $7.5 \times 10^{-14} {\rm eV} \lesssim \mu_1 \lesssim 5.61 \times 10^{-11}  {\rm eV}$ and $2.8 \times 10^{-13}  {\rm eV} \lesssim \mu_2 \lesssim 4.4 \times 10^{-11}  {\rm eV}$ for spin-0,-1 and -2 bosons respectively. In the case of the QCD axion this pushes approximate bounds on sub-Plankian decay constants to values of $f_a \gtrsim 1.4 \times 10^{17}  {\rm GeV}$ for the 2MASS J05215658+4359220 BH and $f_a \gtrsim 1.1 \times 10^{17}  {\rm GeV}$ from the secondary component of GW190814. 

Robust observational evidence of X-ray binaries still suggest a higher bound on the upper mass gap and lower bound on the empirical minimal mass of BHs to conform to $M_{\rm BH}\gtrsim 5M_{\odot}$. The lightest of these, GRO J1655-40 found in Table~\ref{tab:stellarBHs} currently fixes bounds on scalar masses less than $\mu_0 \lesssim 1.6 \times 10^{-11} $ eV or $f_a \gtrsim 3.6 \times 10^{17}$ GeV. A deeper exploration and clarification of the nature of this gap or even its existence along with the evolutionary physics of systems in this region which may contain \emph{hypermassive} NS \cite{Huang:2020cab,Godzieba:2020tjn,Tsokaros:2020hli}, \emph{ultralight} stellar mass BHs and/or PBHs \cite{Carr:2020xqk,Jedamzik:2020omx,Clesse:2020ghq,Vattis:2020iuz} formed in the early Universe due to gravitational collapse of density fluctuations, may push these limits even further. However substantial work on stellar formation theory needs to be done first.

\subsubsection{The Extragalactic Stellar Graveyard, Intermediate Mass Black Holes \& Light Galactic Cores}

From a theoretical standing an absence of specific IMBHs sitting in the BH mass-spin window is well motivated by grand unification arguments in M-theory. The $G_2$-MSSM \cite{friedmann2003,Acharya:2012tw,Acharya:2006ia,Acharya:2007rc,Acharya:2008zi,Acharya:2008hi,Acharya:2010zx,Ellis2015} represents an effective four-dimensional $\mathcal{N}=1$ supergravity theory arising from the KK-compactification of a seven-dimensional singular manifold of $G_2$ holonomy. The phenomenological landscape of these M-theory frameworks concerns the stabilisation of a single class of moduli, the three-form periods over the basis three-cycles of the extra-dimensional manifold. Considering known results for singularities on compact $G_2$ manifolds (i.e. those which generate features such as chiral fermions or well defined local metrics), a key principle component motivating the low-energy effective phenomenology is the volume of three-manifold, $\mathcal{V}_{\rm Vis}$, which supports the the visible sector Minimal Supersymmetric Standard Model (MSSM) gauge group,
\begin{equation}
    \alpha_{\rm GUT} = \frac{g_{\rm GUT}^2}{4\pi} = \frac{\left(4\pi \right)^{\sfrac{1}{3}}g^2_{\rm 11D}}{\mathcal{V}_{\chi}} \ .
\end{equation}
The value $g^2_{\rm 11D}$, represents the fundamental eleven-dimensional coupling related to the eleven-dimensional Planck mass, $M_{\rm 11D}\sim \mathcal{O}(1)\times10^{17}\ {\rm GeV}$ (from moduli stablisation results) via the relationship, $2g^2_{\rm 11D}=\left(2\pi \right)^8 M^{-9}_{11\rm D}$. Standard results of the MSSM enforce $\alpha^{-1}_{\rm GUT} \equiv {\rm IM}(f) \approx 25$ (where $f$ is the visible sector gauge kinetic function) for reconstruction of the visible sector or that the visible sector gauge group is supported on a three-cycle with the approximate volume, $\mathcal{V}_{\rm VIS} \approx 25$. The ability to constrain the mass of the axionic supersymmetric partner of the geometric moduli associated to this cycle, can  indicate values of the gauge unification coupling constant consistent in these models. The dynamical scale of the ultralight axion potential,  generated from higher order corrections to the superpotential such as membrane instantons, can be parameterised in terms of the gravitino mass, $m_{\sfrac{3}{2}}$ up to $\mathcal{O}(1)$ corrections to the exponent, via the expression, $\mu_{a_i} \simeq \mathcal{O}(10^{-3})\sfrac{M_{\rm Pl}}{M_{\rm GUT}}\sqrt{(m_{3/2} M_{\rm Pl} )}e^{-0.5 b_i \mathcal{V}_i}$ \cite{Acharya:2010zx}. 

The window for GUT axions (assuming $f_a \gtrsim 10^{16}$ GeV) is currently bounded by the largest observed stellar mass BHs and the lowest mass observations of low mass AGN. The recent observation of the binary merger GW190521 at redshift 0.8 contained two high mass BHs, the primary component significantly larger than the stellar structure theory limit from PISN. The largest of these ($M_{\rm BH}=85^{+21}_{-14}\ M_{\odot}$) represents the heaviest recorded stellar mass BH to-date, leading to the lightest bounds in the $\mu \sim 10^{-14}$ eV mass range. Although the formation channel of a BH appearing to sit in the PISN mass gap is unclear we make the simplified assumption the components of the binary system are independently long-lived to be sufficiently effected by superradiance. There are expected to be many sources of uncertainty in the nature of this gap such as nuclear reaction rates or the collapse of the hydrogen envelope \cite{Abbott:2020mjq}. It could therefore be that the previous bounds on the higher map gap were conservative lower estimates to the true value. 

On the other side of the exclusion bound dip, an example LMBH candidate with a direct mass and spin constraints is  UGC 06728 \cite{2016ApJ...831....2B,2013MNRAS.428.2901W}, a late-type, low-luminosity Seyfert 1 galaxy. The mass of the BH is determined as $M_{\rm BH} = 7.1 \pm 4.0 \times 10^{5} M_{\odot}$ and its spin limited to $a_*\geq 0.7$. Although this BH has the lowest mass in the total ensemble we consider, the higher spin bounds on NGC 4151 actually lead to higher mass boson constraints, highlighting the spin dependence of these types of bounds. Using the bounds defined in Table~\ref{tab:68percentresults} we define the allowed GUT axion window as, 
\begin{equation}
     1.3 \times 10^{-16} {\rm eV} \lesssim \mu_0^{\rm GUT} \lesssim 3.8 \times 10^{-14} {\rm eV}\ .
     \label{eq:gutaxion}
\end{equation}
Future observations of highly spinning LMBHs will squeeze this limit further. 

We can define an example toy model by fixing both $M_{\rm GUT} = 2\times 10^{16}$ GeV and selecting a cosmologically motivated mass for the gravitino, $m_{\sfrac{3}{2}} =30 $ TeV, based on naturalness arguments and concerns with mass of the lightest modulus \cite{Acharya:2012tw}. We also fix the membrane instanton integers to $b_i = 2\pi$ for concreteness. Using the allowed window for the GUT axion in Eq.~(\ref{eq:gutaxion}), whilst allowing for $\mathcal{O}(1)$ fluctuations (see Eq.~(11) of Ref.~(\cite{Acharya:2010zx})) gives the approximate window limits $(26.4-25.9) \lesssim \mathcal{V}_{\rm GUT} \lesssim (24.6-24.1)$ for the three-dimensional sub-manifold volume of the visible sector or $\sfrac{1}{(25.9-26.4)} \lesssim \alpha_{\rm GUT} \lesssim \sfrac{1}{(24.1-24.6)} $ for the GUT coupling constant. Similarly the current exclusion window for axion masses roughly translates to exclusions on volumes of $29.4 \lesssim \mathcal{V}_{\chi} \lesssim 25.9$ and $24.1 \lesssim \mathcal{V}_{\chi} \lesssim 21.9$ in this simplified toy example.    

Of course the precise values of the quantities discussed above are characterised according to a detailed description of the stabilised internal volume and complete supersymmetric compactification process. The universal nature of superradiance does however offer a neat prospect of limiting geometrical model features in the future for specific subsets of these models where extensive clarification on this front is required before making any robust constraints on model parameters. Similarly in Ref.~\cite{Marsh:2019bjr} they consider a model of the axiverse in which one ultralight field plays the role the QCD axion and one (or more) field(s) in the \emph{1Hz} axion window provides an $\mathcal{O}(1)$ contribution to the required cosmological abundance of DM. In this sense the \emph{1Hz} axion window for DM like fields using Eq.~(\ref{eq:gutaxion}), should have fields oscillating with a frequency inside the band, $0.19\ {\rm Hz}\lesssim \nu_{\phi}\lesssim 57.28 \ {\rm Hz}$. This relates to the approximate allowed visible sector dynamical scale, $\Lambda_{\rm Vis}\sim (0.04-0.61)$ MeV.

\subsubsection{Cosmic Giants and Ultralight Bosonic Dark Matter}
\label{sec:dmbounds}

The heaviest BH recorded to date is associated to the quaser TON 618 \footnotetext{In Ref.~\cite{Zu:2020whs} they utilise the UMBH SDSS J140821.67+025733.2 although the indirect measurements make the possible mass estimates significantly large.} with a predicted mass of $6.6 \times 10^{10}M_{\odot}$ \cite{Shemmer:2004ph}. Other example candidates in this range are Holmberg 15A \cite{2019ApJ...887..195M} ($4 \pm0.8 \times 10^{10}$), IC 1101 \cite{2017MNRAS.471.2321D} ($(4\times10^{10}M_{\odot}$), NGC 4889 ($2.1\pm1.6\times10^{10}$) \cite{2011Natur.480..215M,2012ApJ...756..179M} and NGC 3842 ($9.7\pm^{3.0}_{2.5}\times 10^{9}M_{\odot}$) \cite{2011Natur.480..215M,2012ApJ...756..179M}. A particularly attractive model for the mysterious DM content of the Universe asserts the DM particle is a boson with an ultralight mass of the order $\mu^{FDM} \simeq 10^-{21-22}\ {\rm eV}$, its corresponding de Broglie wavelength comparable to a typical galactic length scale ($\sfrac{mv}{\hbar}\sim$ 0.1.kpc assuming $v\sim 100kms^{-1}$). This model is popular due to its ability to alleviate tensions with numerous small scale issues \cite{Marsh:2013ywa}, and is known as fuzzy DM (FDM) or quantum wave DM \cite{PhysRevLett.85.1158,Schive:2014dra}.

If the reasoning behind the maximum mass of a BH from accretion arguments holds true these BHs represent the potential to probe the lightest masses for bosnic fields. Confirming non-zero spin for these candidates would push the current limits of stringent DM bounds. As a representative example we adopt an optimistic proxy prior for the spin, $a_* = 0.9\pm0.1$ under the assumption these BHs might be found to be spinning with extremal values. In Table~\ref{tab:UMBHcon} we give optimistic bounds on ultralight bosons using this assumption. In the case that an estimate is not placed on the error of the BH mass we use $M^{\rm err}_{\rm BH} = 0.5M_{\rm BH}$. A number of these BHs are able to fully exclude the FDM mass range for Spin-1 fields and a large portion for Spin-0.

\begin{table*}[t]
  \centering
    \caption{Example boson mass bounds from a section of the heaviest UMBHs recorded to date, assuming the BH is observed to possess a high value for its spin. We assume a optimistic proxy prior of $a_*=0.9\pm0.1$ with an instability timescale of $\tau_{\rm Sal}$ only as demonstrative example. In the case of spin-1 fields the heaviest recorded BH to date, TON 618 could potentially exclude fields spanning the full domain of FDM. \\} 
  \resizebox{\textwidth}{!}{%
  \begin{tabular}{@{}c|c|c|c@{}}
 \hline
 \noalign{\vskip 0.1cm}
  \hline
    Black Hole  & Spin-0    & Spin-1 & Spin-2 \\
 \hline
 \noalign{\vskip 0.1cm}
 \hline
    TON 618 &$3.6 \times 10^{-22} {\rm eV} \leq \mu_{0} \leq 4.9 \times 10^{-22} {\rm eV}$ & $9.2 \times 10^{-23} {\rm eV} \leq \mu_{1} \leq 1.0 \times 10^{-21} {\rm eV}$  &$2.0 \times 10^{-22} {\rm eV} \leq \mu_{2} \leq 9.6 \times 10^{-22} {\rm eV}$ \\
     Holmberg 15A &$5.3 \times 10^{-22} {\rm eV} \leq \mu_{0} \leq 9.1 \times 10^{-22} {\rm eV}$ & $1.3 \times 10^{-22} {\rm eV} \leq \mu_{1} \leq 1.9 \times 10^{-21} {\rm eV}$  &$2.7 \times 10^{-22} {\rm eV}\leq \mu_{2} \leq 1.5 \times 10^{-21} {\rm eV}$ \\
      IC 1101 &$5.5 \times 10^{-22} {\rm eV} \leq \mu_{0} \leq 8.2 \times 10^{-22} {\rm eV}$ & $1.4 \times 10^{-22} {\rm eV} \leq \mu_{1} \leq 1.7 \times 10^{-21} {\rm eV}$  &$3.0 \times 10^{-22} {\rm eV}\leq \mu_{2} \leq 1.4 \times 10^{-21} {\rm eV}$ \\
       NGC 4889 &$9.6 \times 10^{-22} {\rm eV} \leq \mu_{0} \leq 1.4 \times 10^{-21} {\rm eV}$ & $2.5 \times 10^{-22} {\rm eV} \leq \mu_{1} \leq 2.9 \times 10^{-21} {\rm eV}$  &$5.5 \times 10^{-22} {\rm eV}\leq \mu_{2} \leq 2.6 \times 10^{-21} {\rm eV}$ \\
        NGC 3842 &$1.9 \times 10^{-21} {\rm eV} \leq \mu_{0} \leq 3.8 \times 10^{-21} {\rm eV}$ & $4.7 \times 10^{-22} {\rm eV} \leq \mu_{1} \leq 1.2 \times 10^{-20} {\rm eV}$  &$9.6 \times 10^{-22} {\rm eV}\leq \mu_{2} \leq 5.4 \times 10^{-21} {\rm eV}$ \\
 \hline
 \noalign{\vskip 0.1cm}
    \bottomrule
  \end{tabular}}
      \label{tab:UMBHcon}
\end{table*}

 Of the current heaviest BHs with Kerr parameter estimations, example candidates which can constrain the lightest possible bosonic fields are M87* \cite{Akiyama:2019cqa,Akiyama:2019brx,Akiyama:2019sww,Akiyama:2019eap,Akiyama:2019fyp} at the center of the super-giant elliptical galaxy Messier 87, quasar H1821+643 and the blazar OJ 287 primary. Previous constraints have been found for M87* in Refs.~\cite{PhysRevLett.123.021102,PhysRevLett.124.061102} and for OJ 287 primary in Ref.~\cite{Zu:2020whs}. The mass of M87* is well defined from the Event Horizon Telescope observations of the BHs shadow \cite{Akiyama:2019fyp}, 
\begin{equation}
    M^{\rm M87^*}_{\rm BH} = 6.5\pm 0.7 \times 10^{9} M_{\odot} \ ,
\end{equation}   
which is consistent with previous mass estimates through the kinematics of gas dynamics \cite{Akiyama:2019eap,1994ApJ...435L..35H,Macchetto_1997,Walsh_2013} ($\sim 3 \times 10^9 M_{\odot}$) and stellar dynamics \cite{Akiyama:2019eap,Gebhardt_2009,Gebhardt_2011} ($\sim 6.5 \times 10^9 M_{\odot}$). Its background metric solution was also shown to be compatible with the Kerr solution \cite{Akiyama:2019cqa}. Its mass may also be closer to heavier UMBHs from considerations for a SSA-thick ISCO ring \cite{Kawashima:2019ljv} ($9\times 10^9 M_\odot$). The spin of the BH is generally expected to be high \cite{2012Sci...338..355D,Takahashi_2018,Nakamura_2018}, although not currently a well defined parameter. Previous results consist of estimations from its twisted light \cite{Tamburini:2019vrf} ($0.9\pm 0.05$), \cite{Dokuchaev:2019pcx} ($0.75\pm 0.15$),  \cite{Bambi:2019tjh} ($a_*<0.95$ for Kerr), TeV photon observations \cite{2009ApJ...699..513L} ($a_*>0.8$), general relativistic magnetohydrodynamic models of jet formation \cite{Nemmen:2019idv} ($|a_*|>0.5$) and observed jet power spectra \cite{Akiyama:2019fyp} ($0.94>|a_*|>0.5$). Other possibilities include M87* as a superspinar \cite{Bambi:2019tjh} ($|a_*|>1$) or low spin Kerr solution (i.e. jet boundary constraints \cite{Nokhrina:2019sxv} ($a_*\sim 0.2-0.3$)). For the purposes of defining optimistic constraints we adopt the prior \cite{Tamburini:2019vrf}, 
\begin{equation}
    a_*^{\rm M87^*} = 0.9 \pm 0.05\ .
\end{equation}
The relevant BH timescale for M87* which concerns its mass accretion rate, $\dot{M}_{\rm BH}$, may in-fact be significantly suppressed compared to the Bondi accretion rate \cite{Kuo:2014pqa,Akiyama:2019fyp}. To conservatively account for this follow Ref.~\cite{PhysRevLett.123.021102} and allow the spin-down to occur within a Hubble time, $\tau_{\rm BH}\sim 10^{10}\ {\rm Yrs}$ which gives the following exclusion windows, 
\begin{align}
    1.3 \times 10^{-21} {\rm eV} &\leq \mu_{0} \leq 1.8 \times 10^{-20} {\rm eV}\ ,  \\
     2.9 \times 10^{-22} {\rm eV} &\leq \mu_{1} \leq 2.6 \times 10^{-20} {\rm eV}\ , \\
      7.2 \times 10^{-22} {\rm eV} &\leq \mu_{2} \leq 2.5 \times 10^{-20} {\rm eV}\ .
\end{align}
In the \emph{right panel} of Fig.~\ref{fig:boseplots} we present 68\% confidence limit bounds bounds for ultralight scalar DM in the $(f_a,\mu_0)$ plane from M87* along with Ark 120 for comparison. The high spin/ high mass of M87* gives the strongest bounds on ultralight spin-0 fields approaching FDM mass scales, giving a strong bound for masses $\mu_0\sim 2.5 \times 10^{-21}\ {\rm eV}$ with $f_a \gtrsim 10^{16}\ {\rm GeV} \approx M_{\rm GUT}$, consistent with previous results \cite{PhysRevLett.123.021102}. The lower spin values and large uncertainty of H1821+643 only returns bounds for spin-1 fields over a characteristic timescale of $\tau_{\rm Hub}$, $5.3 \times 10^{-22} {\rm eV} \leq \mu_{1} \leq 8.8 \times 10^{-22} {\rm eV} $. For OJ 287 primary the spin of the BH is sufficiently low, superradiance does not exclude spin-0 fields over timescales $\tau_{\rm BH}< \tau_{\rm Hub}$ and therefore we do not find bounds on the self-interaction strength either for these two BHs. For spin-1 and spin-2 bosons with $\tau_{\rm BH}< \tau_{\rm Hub}$ we find the exclusion windows $2.9 \times 10^{-22} {\rm eV} \leq \mu_{1} \leq 7.2 \times 10^{-22} {\rm eV} $ and $6.7 \times 10^{-22} {\rm eV} \leq \mu_{2} \leq 1.4 \times 10^{-21} {\rm eV} $ respectively. These bounds represent constraints on lightest bosonic matter from BH spin measurements whilst also limiting the masses of potential candidates in the FDM scenario.

\section{Multiple Bosonic Fields in the Ultralight dark sector}
\label{sec:multifield}

Given the broad logarithmic spread of the current bounds on ultralight bosonic fields from superradiance, it is also natural to consider features of certain model spectra populating the limits given in Section~\ref{sec:frefieldresults} \cite{Stott:2018opm,Stott:2018axz}. The string axiverse scenario \cite{Arvanitaki:2009fg,Arvanitaki:2010sy,Cicoli:2012sz} often predicts the appearance of $\mathcal{O}(10)\text{-} \mathcal{O}(100)$ scalar degrees of freedom in the effective limits of superstring models. A common phenomenological prior placed on the masses of these fields is a flat prior, motivated by the topological complexity of the extra-dimensional manifold and the exponential dependence the fields mass to the action of their corresponding cycles \cite{Svrcek:2006yi,Svrcek:2006hf}. This \emph{maximally ignorant} baseline approach indicates these fields may extend from the Planck scale down to the Hubble scale today ($\sim 10^{-33}\ {\rm eV}$) \cite{Kamionkowski:2014zda,Stott:2017hvl,Visinelli:2018utg}. It has also been shown the mass spectrum associated to ultralight fields in the dark sector can be modeled using a high-dimensional RMT analysis \cite{Stott:2017hvl,Stott:2018opm}. This approach invokes unimodal measures on the mass eigenstates in the limit of a large population of fields, which take the form of well defined limiting spectral distributions. In the simplest configurations the nature of the spectrum is principally regulated by a two hyperparameters $\zeta = \{\beta,\bar{\mu}^2\}$. The value $\beta = \sfrac{\mathscr{N}}{\mathscr{P}}$ is determined by the dimensionalty of the sample matrices, with $\mathscr{N}$ (number of fields) samples and $\mathscr{P}$ (unknown) variables. The value of $\bar{\mu}$ fixes the mean of the dimensionful scales in the model. 
\begin{figure*}[!tbp]
  \centering
  \begin{minipage}[b]{0.49\textwidth}
    \includegraphics[width=\textwidth]{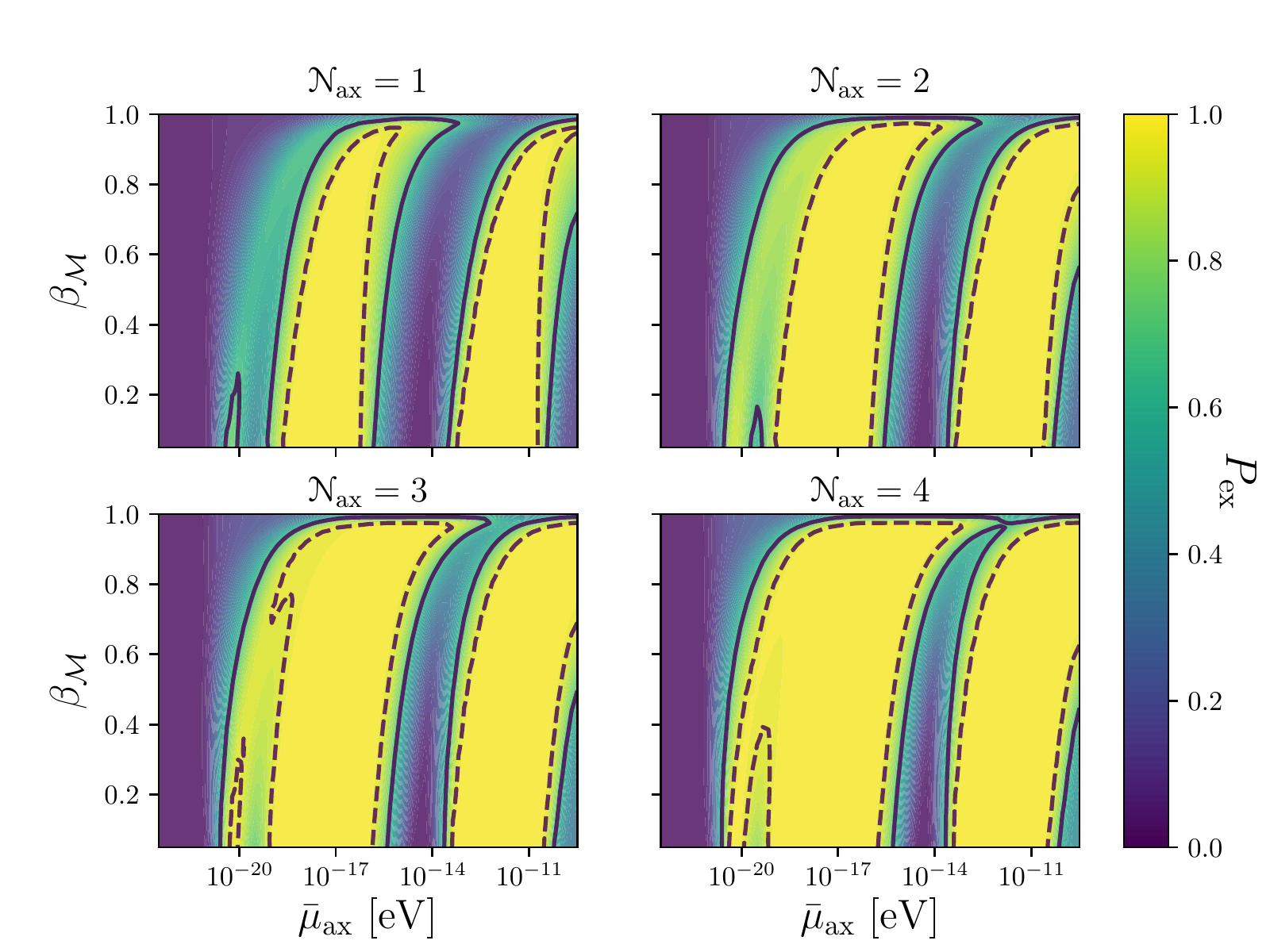}
  \end{minipage}
  \hfill
  \begin{minipage}[b]{0.49\textwidth}
    \includegraphics[width=\textwidth]{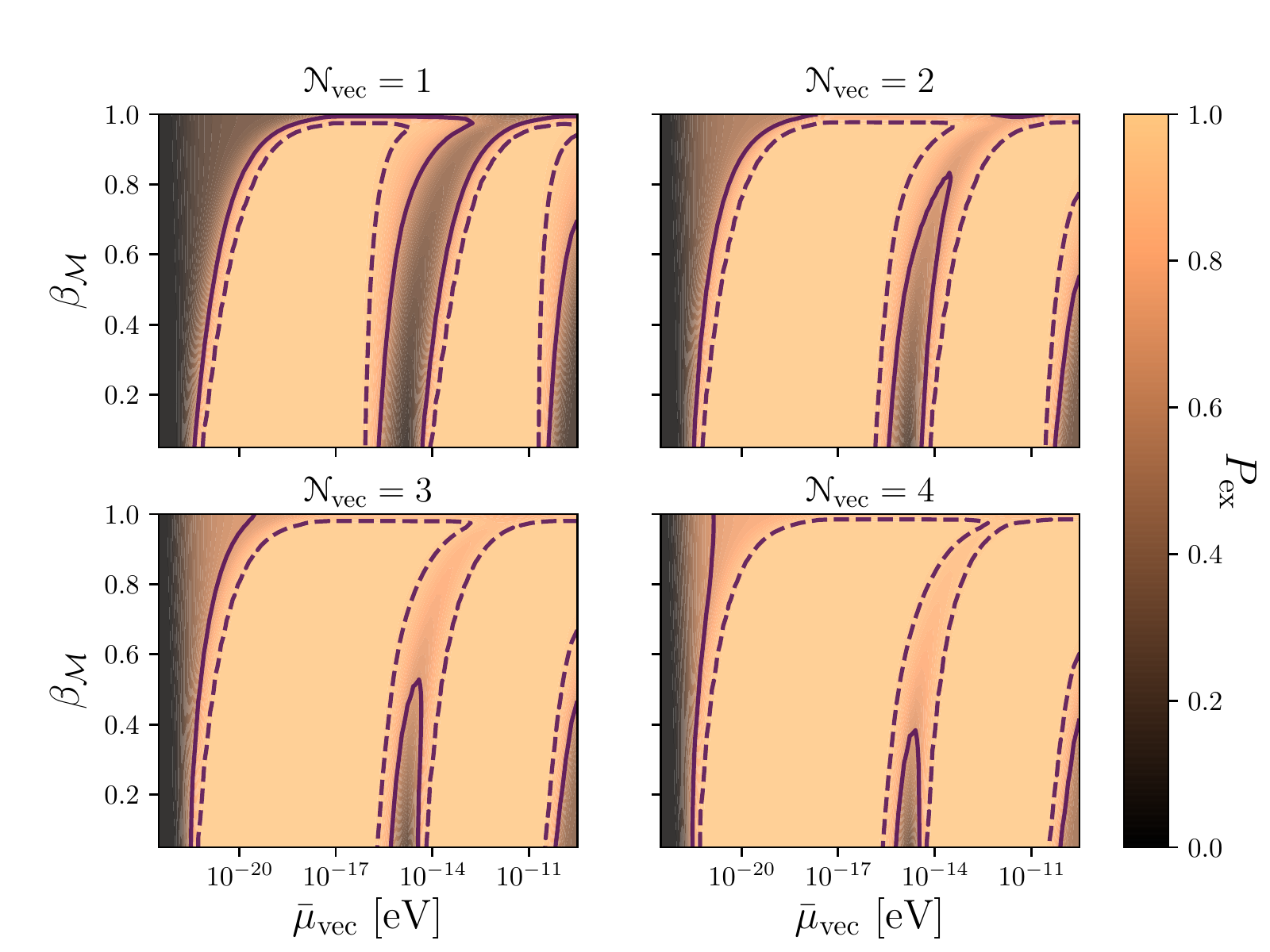}
  \end{minipage}
  \caption{Example bounds for massive spin-0 and spin-1 fields drawn according to the mass spectrum discussed in Appendix~\ref{app:multifields}. Each spectrum is parameterised according to a mean scale denoted, $\bar{\mu}$ and a distribution shaping parameter, $\beta_{\mathcal{M}}$. We calculate the bounds using Eq.~(\ref{eq:multifieldeq}) with the functions presented in Fig~\ref{fig:p_explot} and the spectrum support in Eq.~(\ref{eq:mpeq}). In each panel the \emph{solid black} line represents the 68\% confidence isocontour and the \emph{dashed black} line represents the 95\% confidence isocontour. In the \emph{left panel} we present the bounds for spin-0 fields for the cases of $\mathscr{N}_{\rm ax}=1$ to $\mathscr{N}_{\rm ax}=4$ using the effective model in Appedix~\ref{app:spin0multi}. In the \emph{right panel} we present the bounds for spin-1 fields for the cases of $\mathscr{N}_{\rm vec}=1$ to $\mathscr{N}_{\rm vec}=4$ using the effective model in Appedix~\ref{app:spin1multi}.}
  \label{fig:multiconst}
\end{figure*}

To consider bounds using an example spectrum we adopt the methodology of Ref.~\cite{Stott:2018opm} and the effective field theory given detailed in Appendix~\ref{app:multifields} and discussed in Refs.~\cite{Stott:2017hvl,Stott:2018axz}. We assume the presence of a sufficient number of fields ($\mathscr{N}\gtrsim \mathcal{O}(5-10)$) in the total spectrum such that the limiting spectral prior converges to the form in Eq.~(\ref{eq:mpeq}) from universality arguments. In the case of massive spin-0 fields (Appendix~\ref{app:spin0multi}) previous analysis have considered the spectrum of the mass matrix well-approximated by a Wishart matrix \cite{Bachlechner:2014hsa,Battefeld:2008bu,Baumann:2014nda,Braden:2010wd,Christodoulidis:2019hhq,Cicoli:2014sva,Easther:2005zr,Easther:2013rva,Kim:2007bc,Price:2014xpa,Price:2015qqb,Stott:2017hvl} as well as analyses for non-trivial kinetic matrices and charge matrices exploring the fundamental domain of the axion field space using canonical ensembles of positive definite random matrices \cite{Bachlechner:2015gwa,Bachlechner:2014hsa,Bachlechner:2014gfa,Bachlechner:2015qja,Bachlechner:2017hsj,PhysRevD.98.061301,Bachlechner:2018gew,Bachlechner:2019vcb,Ferrari:2011is,Heidenreich:2019bjd}. For massive vectors (Appendix~\ref{app:spin1multi}) we explore a mirrored extension to the spin-0 case as a comparative example but offer no physical motivations. Using the mass exclusion function determined from superradiance we can exclude a defined prior for the mass spectrum via an evaluation of a simple numerical product integral for a set of hyperparameters $\zeta$ of a model $\mathscr{M}$ \cite{Stott:2018opm}, 
 \begin{equation}
     P_{\rm al}(\zeta,\mathscr{N}_s|\mathscr{M}) = \left[ \int d \mu_{s} P(\mu_s,\zeta|\mathscr{M})P_{\rm al}(\mu_s,\mathscr{N}_s=1) \right]^{\mathscr{N}_s}\ ,
     \label{eq:multifieldeq}
 \end{equation}
where $P_{\rm al}(\mu_s,\mathscr{N}_{s}=1) = 1 - P_{\rm ex}(\mu_s,\mathscr{N}_{s}=1)$ and $P_{\rm ex}(\mu_s,\mathscr{N}_{s}=1)$ is one of the functions visualised in Fig.~\ref{fig:p_explot} for spin-0 ($s=0$) or spin-1 ($s=1$) bosons. The function $P(\mu_s,\zeta|\mathscr{M})$ is the spectral prior we fix for the boson masses. If $\mathscr{M}$ represents the limiting spectral distribution of a RMT model then statistically the eigenvalues are correlated quantities and the joint eigenvalue density must be corrected each draw to account for local eigenvalue probabilities. Specifically for individual eigenvalue statistics for matrices residing in canonical invariant ensembles this corresponds to a calculation of spectral quantities such as gap probabilities. These are often expressed in terms of well defined generating kernels used to calculate the relevant $\mathscr{N}$-point correlation functions which define the correct eigenvalue probability density \cite{rmtbook,2004math.ph..11075D,2007JSP...129..949D}. To account for this and avoid such technical complexities we limit the number of fields we `draw' from the limiting spectrum to a maximum of a `few' ($\mathscr{N}_{\rm ax} \leq 4$) in this heuristical example. This allows us the approximation that the eigenvalues are separated \emph{globally} over the compact interval of their spectral support, where localised eigenvalue repulsions from spectral correlations are sufficiently suppressed. 

In the \emph{left panel} of Fig~\ref{fig:multiconst} we present constraints for a spectrum of spin-0 fields determined by the Mar\v{c}henko-Pastur density function (see Appendix~\ref{app:spin0multi}) for different values of its spectral mean ($\bar{\mu}_{\rm ax}$) and distribution shaping parameter ($\beta_{\mathcal{M}}$). In the case of $\mathscr{N}_{\rm ax}=1$ the faded constrained region at the order $\bar{\mu}_{\rm ax} \simeq 10^{-20}\ {\rm eV}$ corresponds to constraints from BHs approaching the UMBH mass range from M87*. As we increase the number of fields these bounds quickly converge into the remaining collective SMBH bounds. Likewise mean spectral scales in the IMBH range also begin to be more heavily constrained. When $\beta_{\mathcal{M}}\rightarrow 0$ the spectrum approaches a point measure $P(\mu_{\rm ax})\simeq \delta(\mu_{\rm ax})$, up to statistical fluctuations and the bounds approach the free-field limits given in Section~\ref{sec:frefieldresults}. Alternatively as $\beta_{\mathcal{M}}\rightarrow 1$ the spectrum support is greatly enhanced over logaritimic distances (see the \emph{right panel} of Fig.~7 in Ref.~\cite{Stott:2018opm}) leading to the degeneracy curves in the constraints which drop out as $\beta_{\mathcal{M}}\simeq 1$. 

Similarly in the \emph{right panel} of Fig.~\ref{fig:multiconst} we present constraints for a spectrum of spin-1 fields also determined by the Mar\v{c}henko-Pastur spectral function (see Appendix~\ref{app:spin1multi}).  The enhanced range of the mass bounds for spin-1 fields (see Table~(\ref{tab:68percentresults})), generates stronger bounds in the multifield case compared to the spin-0 example, where at $\mathscr{N}_{\rm vec} = 4$ spectra with a fixed mean inside $10^{-20}\lesssim \bar{\mu}_{\rm vec} \lesssim 10^{-10}$ are nearly fully constrained at the 68\% confidence limit across the full window. This representative of the characteristic dependence that the phenomena of superradiance partnered with only a small number of fields in a spectrum located within this mass range can quickly constrain the spectral hyperparameters of models \cite{Stott:2018opm}. Accounting for eigenvalues correlations would allow for bounds in cases where the mean scale is fixed far away from the limits considered in Fig.~\ref{fig:multiconst}, as a larger number of fields will enhance the likelihoods of drawing outliers which can still provide exclusions.

\section{Conclusion}
\label{sec:conclusion}

Models of ultralight bosonic fields offer numerous enticing possibilities but come with the logistical challenge of detection induced by their extremely weak couplings. Superradiant instabilities offer a unique opportunity to place bounds on the weakly-coupled sector of ultralight massive bosons through indirect methods. A statistical analysis of the Regge mass-spin plane with measurements of astrophysical BHs with non-zero spin can be used to exclude specific mass scales and self-interaction strengths. 

In this paper we have considered constraints on ultralight bosonic fields of integer spin-0, -1 and -2 from the parameter measurements of a generous number of identified BHs. Generally our results hold in the limit of a linearised analysis of bosonic instabilities where the cloud is free to evolve independently of features of the surrounding spacetime. Dominant modes are free to spin down the BH without accounting for possible non-linear features such as level-mixing. The bounds on bosons with non-zero spin are more stringent due to the strength of the instability. For spin-2 bosons future clarification on the nature of the complete eigenmode spectrum will provide more accurate bounds \cite{Brito:2013wya,Brito:2015oca,PhysRevLett.124.211101}. Our results for the current exclusion windows using the analytical results discussed in Section~\ref{sec:freefielddomain} for the instability rates, at the 68\% confidence limit (see Table~\ref{tab:95percentresults} for the 95\% bounds) as as follows. For spin-0 bosons: 
\begin{align}
 3.8\times10^{-14}\ {\rm eV} \leq &\mu_0 \leq 3.4\times10^{-11}\ {\rm eV}\ , \\ 
  5.5\times10^{-20}\ {\rm eV} \leq &\mu_0 \leq 1.3\times10^{-16}\ {\rm eV}\ , \\  
   2.5\times10^{-21}\ {\rm eV} \leq &\mu_0 \leq 1.2\times10^{-20}\ {\rm eV}\ . 
\end{align}
For spin-1 bosons:
\begin{align}
    6.2\times10^{-15}\ {\rm eV} \leq &\mu_1 \leq 3.9\times10^{-11}\ {\rm eV}\ , \\  
   2.8\times10^{-22}\ {\rm eV} \leq &\mu_1 \leq 1.9\times10^{-16}\ {\rm eV}\ .   
\end{align}
For spin-2 bosons: 
\begin{align}
  2.2\times10^{-14}\ {\rm eV} \leq &\mu_2 \leq 2.8\times10^{-11}\ {\rm eV}\ , \\ 
  1.8\times10^{-20}\ {\rm eV} \leq &\mu_2 \leq 1.8\times10^{-16}\ {\rm eV}\ , \\  
   6.4\times10^{-22}\ {\rm eV} \leq &\mu_2 \leq 7.7\times10^{-21}\ {\rm eV}\ .    
\end{align}
Incorporated into these exclusion windows are recent observations such as the event GW190521 and the shadow of M87*, each extending the previous bounds in Ref.~\cite{Stott:2018axz}. It is also expected future detections (LISA etc.) of IMBHs will bridge the current gaps, in the exclusion windows above \cite{Brito:2017wnc,Brito:2017zvb}. We have also considered recent interesting observations of BHs sitting above/below mass gaps in the BH mass spectrum which could potentially offer interesting bounds for specific phenomenological models. In particular superradiance can be used to tighten possible window of allowed axion masses in M-theory models, related to supporting the visible section (Eq.~\ref{eq:gutaxion}). We have also explored limited windows for the QCD axion and models of ultralight DM.  

Our analysis has made several key assumptions for simplicity. In particular the systematics of BH spin measurements often lead to poor estimates for these quantities where we have chosen to use a large number of BHs both collectively and individually to place bounds on masses. The most conservative bounds will come from an individual assessment of these systems evolutionary features such as their binary lifetime or accretion rates. It would therefore be interesting to include more direct observations of these BH features into our calculations, see for example Refs~\cite{Cardoso:2018tly,PhysRevD.91.084011}. It may also be possible violent events or specific dynamical features may disrupt the evolution of the cloud in its recent history. To minimally account for this somewhat we have given the individual exclusion bounds for each BH for each type of boson for three characteristic timescales in Appendix~\ref{app:extendedfree}. 

If self-interactions or non-linear effects are to be accounted for this may also effect the growth of the cloud, such as those discussed in Section~\ref{sec:bosenova}. These are natural extension to the work conducted here in order to place bounds on individual systems and spin-2 fields in particular, along with other possibilities such as incorporating a full numerical analysis etc. In Section~\ref{sec:interactingresults} we placed bounds on the self-coupling strength of general spin-0 fields using our complete BH data set. We also displayed the full bounds in the ($f_a,\mu_0$) plane for recent observations of both the primary and secondary components of GW190521 and M87* in Fig.~\ref{fig:boseplots}. 

 For models containing more than one field, as is very typical in string theory for example, strong exclusions can occur if a single field is susceptible to the bounds above. This can be used to place limits on the shapes and mean scales of model spectra, as we demonstrated using toy configurations in Section.~\ref{sec:multifield}. Very generally in examples where bosonic masses follow statistical distributions independent of microscopic quantities we have seen only a very small number of fields are required to exclude spectra in a large portion of the ultralight parameter space. Development of the theoretical aspects surrounding superradiant instabilities, partnered with the enhanced observational reach and accuracy of future experimentation, presents a fascinating and dynamical sector for fundamental physics, with a strong possibility of robust constraints on bosonic fields coming in the age of GW and precision BH astronomy.

\appendix

\section{Black Hole Ensemble Data Set}
\label{app:bhdata}
\begin{table}
   \caption{Mass and dimensionless spin parameter measurements for a selection of X-ray binary systems. Both the mass and spin values are quoted up to $1\sigma$ confidence unless otherwise stated. For a review of stellar mass BHs from X-ray sources see Refs.~\cite{Middleton:2015osa,Miller:2014aaa}.}
\addtocounter{table}{-1}
    \begin{tabularx}{\columnwidth}{p{3cm} p{1.7cm}p{1.7cm}p{1.7cm}}
        \hline
       \text{X-Ray Binary}    & $M^{(1)}_{\rm BH}[M_{\odot}]$        & $a_*^{(1)}$  &     ${\rm Refs.}$       \\
        \hline
GRO J1655-40  & $6.30^{+0.5}_{-0.5}$ & $0.7^{+0.1}_{-0.1}$ & \cite{1538-4357-636-2-L113}/\cite{Greene:2001wd}  \\
GRS 1716-249 & $6.45^{+1.55}_{-1.55}$ & $\geq 0.92$  & \cite{Tao:2019yhu}/\cite{Tao:2019yhu}  \\ 
A 0620-00   & $6.61^{+0.25}_{-0.25}$ & $0.12^{+0.19}_{-0.19}$ & \cite{2010ApJ...710.1127C}/\cite{2010ApJ...718L.122G}  \\
LMC X-3  & $6.98^{+0.56}_{-0.56}$ & $0.25^{+0.13}_{-0.16}$ & \cite{2014ApJ...794..154O}/\cite{Steiner:2014zha}  \\
XTE J1550-564  & $9.10^{+0.61}_{-0.61}$ &$0.34^{+0.37}_{-0.34}$ & \cite{2011ApJ...730...75O}/\cite{2011MNRAS.416..941S}  \\
4U 1543-475  & $9.40^{+1.0}_{-1.0}$ & $0.8^{+0.1}_{-0.1}$ & \cite{Orosz67}/\cite{Shafee:2005ef} \\ 
LMC X-1 & $10.91^{+1.41}_{-1.41}$ & $0.92^{+0.05}_{-0.07}$ & \cite{Orosz:2008kk}/\cite{2009ApJ...701.1076G}  \\  
GRS 1915+105 & $10.10^{+0.6}_{-0.6}$ & $\geq 0.95$ & \cite{Steeghs:2013ksa}/\cite{McClintock:2006xd}  \\
GRS 1124-683 & $11.0^{+1.4}_{-1.4}$ & $0.63^{+0.16}_{-0.16}$ & \cite{Wu_2016}/\cite{Chen_2016} \\
Cygnus X-1 & $14.80^{+1.0}_{-1.0}$ & $\geq 0.983$ & \cite{2011ApJ...742...84O}/\cite{Gou:2013dna} \\
M33 X-7 & $15.65^{+1.45}_{-1.45}$ &$0.84^{+0.05}_{-0.05}$ &\cite{Orosz:2007ng}/\cite{1538-4357-679-1-L37} \\
        \hline
        \label{tab:stellarBHs}
    \end{tabularx}
\end{table}
In order to constrain bosonic masses we utilise a generous ensemble of both stellar mass and SMBHs with both parameter estimations and error bounds for their masses and spins. We factorise each into one of three groups according to their mass or observational signature. Until recently most measurements have been gathered using data from accretion observations in AGN or X-ray binary systems, taking advantage of well defined techniques such as thermal continuum fitting \cite{McClintock:2013vwa} or X-ray reflection spectroscopy \cite{Miller:2014aaa,Reynolds:2013qqa,Reynolds:2019uxi}. Mass and spin measurements of known SMBHs residing in AGN are given in Table~\ref{tab:SMBH}. A selection of X-ray binary system \cite{Remillard:2006fc,Miller:2014aaa} measurements for stellar mass sources, which generally have well defined parameters due to their vicinity in the sky is given in Table~\ref{tab:stellarBHs}.

\begin{table}[t]
    \caption{Mass and dimensionless spin parameters measurements for each BBH merger detailed in both the O1 and O2 runs for LIGO along with additional initial results from the O3 run. The data from the O1/O2 runs is detailed in Ref.~\cite{LIGOScientific:2018mvr}. The O3 run data (GW190412, GW190814 and GW190521) can be found in Ref.~\cite{Abbott:2020khf}, Ref.~\cite{LIGOScientific:2020stg} and Ref.~\cite{PhysRevLett.125.101102} respectively. }
    \label{tab:ligodata}
\addtocounter{table}{-1}
    \begin{tabularx}{\columnwidth}{p{2cm} p{1.5cm}p{1.5cm}p{1.5cm}p{1.5cm}}
        \hline
       Event                & $M^{(1)}_{\rm BH}[M_{\odot}]$ & $a_*^{(1)}$  & $M^{(2)}_{\rm BH}[M_{\odot}]$ & $a_*^{(2)}$   \\
        \hline
GW150914  & $35.6^{+4.7}_{-3.1}$ & $0.28^{+0.57}_{-0.25}$ & $30.6^{+3.0}_{-4.4}$ & $0.34^{+0.53}_{-0.30}$ \\
GW151012   & $23.2^{+14.9}_{-5.5}$ & $0.33^{+0.54}_{-0.29}$ & $13.6^{+4.1}_{-4.8}$ & $0.45^{+0.48}_{-0.40}$ \\
GW151226  & $13.7^{+8.8}_{-3.2}$ & $0.57^{+0.36}_{-0.43}$ & $7.7^{+2.2}_{-2.5}$ & $0.51^{+0.44}_{-0.45}$ \\
GW170104  & $30.8^{+7.3}_{-5.6}$ & $0.34^{+0.52}_{-0.30}$ & $20.0^{+4.9}_{-4.6}$ & $0.43^{+0.48}_{-0.38}$ \\
GW170608  & $11.0^{+5.5}_{-1.7}$ & $0.32^{+0.50}_{-0.28}$ & $7.6^{+1.4}_{-2.2}$ & $0.40^{+0.52}_{-0.36}$ \\ 
GW170729 & $50.2^{+16.2}_{-10.2}$ & $0.69^{+0.28}_{-0.55}$ & $34.0^{+9.1}_{-10.1}$ & $0.55^{+0.40}_{-0.49}$ \\    
GW170809 & $35.0^{+8.3}_{-5.9}$ & $0.32^{+0.53}_{-0.29}$ & $23.8^{+5.1}_{-5.2}$ & $0.42^{+0.50}_{-0.37}$ \\
GW170814 & $30.6^{+5.6}_{-3.0}$ & $0.40^{+0.52}_{-0.36}$ & $25.2^{+2.8}_{-4.0}$ & $0.42^{+0.51}_{-0.37}$ \\
GW170818 & $35.4^{+7.5}_{-4.7}$ & $0.46^{+0.48}_{-0.41}$ & $26.7^{+4.3}_{-5.2}$ & $0.46^{+0.47}_{-0.42}$ \\
GW170823 & $39.5^{+11.2}_{-6.7}$ & $0.42^{+0.49}_{-0.37}$ & $29.0^{+6.7}_{-7.8}$ & $0.45^{+0.48}_{-0.40}$ \\
GW190412 & $29.7^{+5.0}_{-5.3}$ & $0.43^{+0.16}_{-0.26}$ &\ \ \ \ - & \ \ \ \  \ \ -\\
GW190814 & $23.2^{+1.1}_{-1.0}$ & $\leq 0.07$ & \ \ \ \ - & \ \ \ \  \ \ -\\
GW190521 & $85.0^{+21.0}_{-14.0}$ & $0.69^{+0.27}_{-0.62}$ & $66.0^{+17.0}_{-18.0}$ & $0.73^{+0.24}_{-0.64}$ \\
\hline
    \end{tabularx}
\end{table}

The new age of GW astronomy \cite{Abbott:2016blz} has already presented us with substantial initial data from LIGOs first three runs (O1/O2/O3). In Table~\ref{tab:ligodata}\footnote{See Ref.~\cite{Zackay:2019btq,Nitz:2019hdf} for discussions of sources which have lower probabilities they are astrophysical in origin but may give indications of further events transcending the upper stellar mass gap. }
    \footnote{See Refs.~\cite{Zevin_2020,Mandel:2020lhv} for an alternative interpretation of this event along with spin estimation for the secondary BH candidate.}
    \footnote{See also Refs.~\cite{Zackay:2019tzo,Venumadhav:2019lyq,Pratten:2020ruz} for possible events not considered in our sample data set (GW170121, GW170304, GW170727).}we give the current complete set of BBH merger events with determined mass and spin bound measurements. Observations of the inspiral, coalescence and ringdown of these BBH systems are used to determine mass and spin estimates of each of the primary and secondary components, both before the merger event. Conservatively one cannot generally make use of the remnant system measurements as $\tau_{\rm Obs} \ll \tau_{\rm SR}$ but this can offer significance with certain configurations \cite{PhysRevLett.124.211101}. Such events are often affected by large errors, but do offer observed examples of BH masses approaching \cite{LIGOScientific:2018jsj,Chatziioannou:2019dsz} and breaching \cite{Abbott:2020tfl,Abbott:2020mjq} the limit \cite{Fishbach_2017,Farmer:2019jed} arising from PISN and stellar formation theory  \cite{1964ApJS....9..201F,PhysRevLett.18.379,Woosley_2017}. Future experiments are expected to significantly reduce these uncertainties for individual spin measurements \cite{TheLIGOScientific:2016pea,Salemi:2019owp}, whilst proving tighter constraints on the upper and lower mass limits for stellar mass systems \cite{Fishbach_2017}. This along with a thorough  understanding of the history and dynamics of these observations will significantly strengthen the ability to place robust bounds on ultralight bosons.

\begin{table}
    \caption{Mass and dimensionless spin parameter measurements for a selection of SMBHs. Mass values are quoted up to $1\sigma$ confidence, with the spins quoted at the 90\% confidence level. For a review of AGN data see Refs.~\cite{2011ApJ...736..103B,Reynolds:2013qqa,Reynolds:2013rva}.}
    \addtocounter{table}{-1}
    \begin{tabularx}{\columnwidth}{p{2.5cm} p{2.2cm}p{1.7cm}p{1.7cm}}
        \hline
      \text{AGN}    & $M^{(1)}_{\rm BH}[10^{6}M_{\odot}]$        & $a_*^{(1)}$  &     ${\rm Refs.}$    \\
        \hline
    Fairall 9  & $255.0^{+56.0}_{-56.0}$ & $0.52^{+0.19}_{-0.15}$ & \cite{Peterson:2004nu}/\cite{2012ApJ...758...67L}   \\
Mrk 79   & $52.40^{+14.40}_{-14.40}$ & $0.70^{+0.1}_{-0.1}$ & \cite{Peterson:2004nu}/\cite{2011MNRAS.411..607G} \\
NGC 3783   & $29.80^{+5.40}_{-5.40}$ & $\geq 0.98$ & \cite{Peterson:2004nu}/\cite{2011Brenneman}  \\
Mrk 335  & $14.20^{+3.70}_{-3.70}$ & $0.83^{+0.09}_{-0.13}$ & \cite{Peterson:2004nu}/\cite{2013MNRAS.428.2901W} \\
MCG-6-30-15   & $2.90^{+1.80}_{-1.60}$ & $\geq 0.98$ & \cite{McHardy:2005ut}/\cite{Brenneman:2006hw}  \\ 
Mrk 110 & $25.10^{+6.10}_{-6.10}$ & $\geq 0.89$ & \cite{Peterson:2004nu}/\cite{2013MNRAS.428.2901W}  \\    
NGC 7469 & $12.20^{+1.40}_{-1.40}$  & $0.69^{+0.09}_{-0.09}$ & \cite{Peterson:2004nu}/\cite{doi:10.1111/j.1365-2966.2011.19224.x}   \\
Ark 120 & $150.0^{+19.0}_{-19.0}$ & $0.64^{+0.19}_{-0.11}$ & \cite{Peterson:2004nu}/\cite{2013MNRAS.428.2901W}  \\
3C120 & $55.0^{+31.0}_{-22.0}$ & $\geq 0.95$  &\cite{Peterson:2004nu}/\cite{Lohfink:2013uwa}  \\
ESO 362-G18 & $12.5^{+4.5}_{-4.5}$ & $\geq 0.92$ & \cite{Agis-Gonzalez:2014nja}/\cite{Agis-Gonzalez:2014nja}  \\
H1821+643  & $4500.0^{+1500.0}_{-1500.0}$ & $\geq 0.4$ & \cite{Reynolds_2014}/\cite{Reynolds_2014}  \\
NGC 4051 & $1.91^{+0.78}_{-0.78}$ & $\geq 0.99$ & \cite{Peterson:2004nu}/\cite{8175999}  \\
NGC 4151 & $45.7^{+5.70}_{-4.70}$ & $\geq 0.9$ & \cite{Bentz:2006ks}/\cite{Keck:2015iqa}  \\
M87* & $6500.0^{+700.0}_{-700.0}$ & $0.9^{+0.1}_{-0.1}$ &\cite{Akiyama:2019fyp}/\cite{Tamburini:2019vrf} \\
OJ 287  & $18348.0^{+7.92}_{-7.92}$ & $0.381^{+0.004}_{-0.004}$ &\cite{Dey:2018mjg}/\cite{Dey:2018mjg} \\
UGC 06728  & $7.10^{+4.00}_{-4.00}$ & $\geq 0.7$ &\cite{2016ApJ...831....2B}/\cite{2013MNRAS.428.2901W} \\
 \hline
 \label{tab:SMBH}
    \end{tabularx}
    
\end{table}

\section{Signals in the Regge Plane}
\label{app:signals}
An additional non-linear component for isolated systems, to the presence of the bosenova discussed in Section~\ref{sec:bosenova} during the superradiant evolution of the bosonic cloud is the possibility of perturbations and level mixing \cite{Arvanitaki:2010sy,PhysRevD.91.084011}. Assuming a hypothetical BH forms in an astrophysical process in the \emph{left panel} of Fig.~\ref{fig:regges}, with a spin greater than the $l=2$ (\emph{dashed orange line}) but less than $l=m=1$ state Reggee trajectory (\emph{dashed blue line}) then the superradiance process for some fixed boson mass will extract angular momentum from the BH until it sits on the $l=2$ line. At this point non-linear self-interactions delay the exponential growth of the subsequent $l=3$ state, through perturbations of the $l=2$ state and potential surrounding the BH, with modes not satisfying the superradiance condition. This level mixing effect must be quenched through boson-graviton annihilations which deplete the cloud for the $l=2$ state. This timescale defines the time the BH spends on a Regge tragectory, 
\begin{equation}
    \tau_{\rm Regge} \simeq \frac{\sqrt{|\sfrac{|\Gamma_{\rm SR}({l-1}}{\Gamma_{\rm SR}^{l+1}}|}}{N_{\rm Bose}\Gamma_{\rm ann}} \ , 
\end{equation}
where $N_{\rm Bose}$ is defined in Eq.~(\ref{eq:nmax}) and $\Gamma_{\rm ann}$ represents the annihilation rate of the bosons to gravitons (see Eq.~(24) of Ref.~\cite{Arvanitaki:2016qwi}). In the case of strongly interacting fields, corresponding to lower values of $f_a$, it is possible that $\tau_{\rm Regge}\gg \tau_{\rm SR}$. Mapping a population of BHs stuck across Regge trajectories therefore indicates to existence of a massive boson.  
\begin{figure}[t]
    \includegraphics[width=0.49\textwidth]{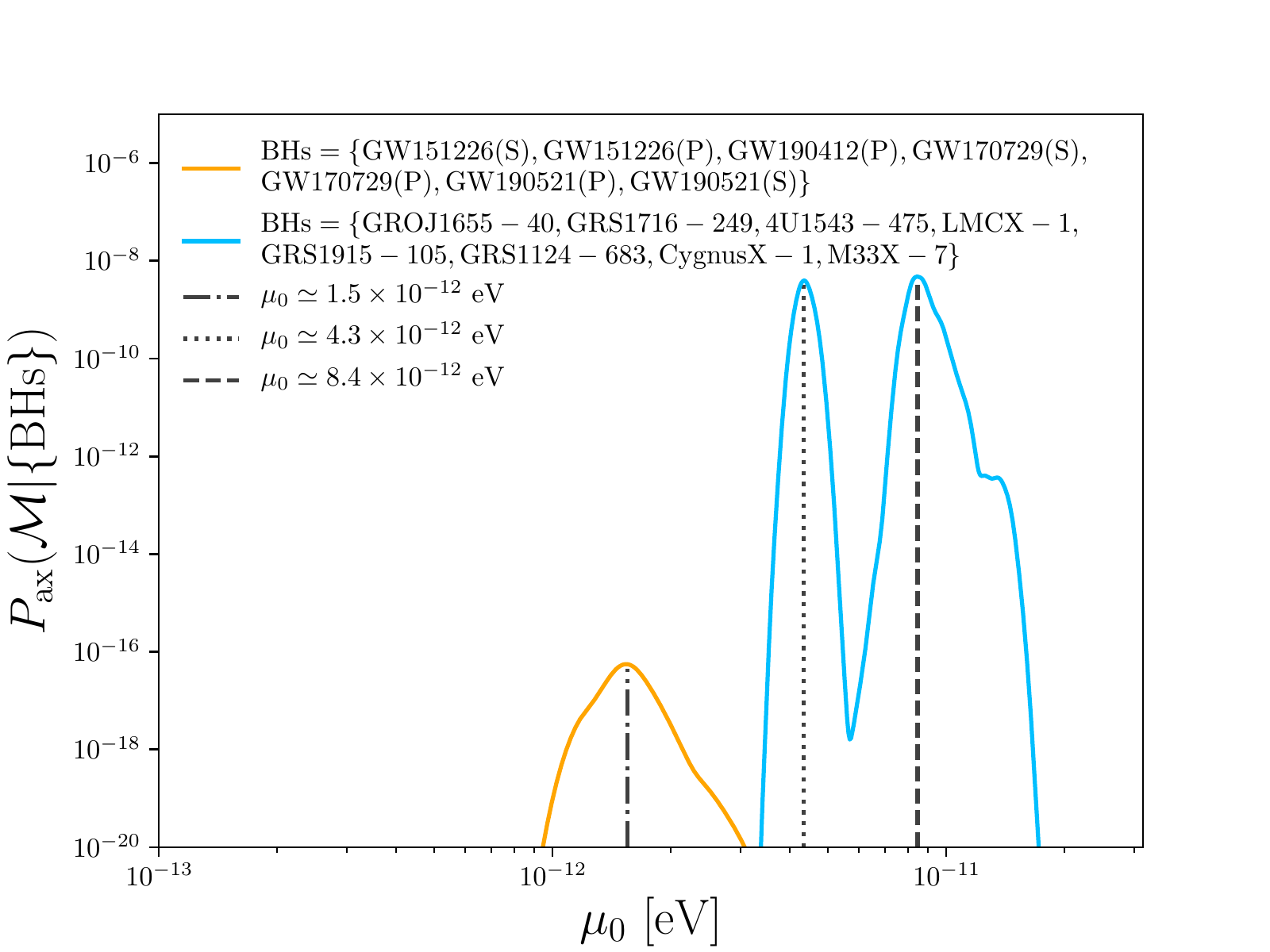}
    \caption{Proability of BH data subsets aligning to Regge trajectories as a function of a scalar boson mass. The \emph{blue} line represents the statistical probability a subset of binary X-ray sources with spins $a_*>0.5$ indicate the presence of a ultralight scalar boson at fixed mass. The \emph{orange} line represents the statistical probability a subset of binary BH merger sources with spins $a_* > \pm a_*^{\rm max}$ indicate the presence of a ultralight scalar boson.}
    \label{fig:regge_results}
\end{figure}
 
To date the large statistical inaccuracies of BH spin measurements leave it difficult to fit mass-spin data to the superradiance condition boundaries (Eq.~(\ref{eq:condition})) for each mode which may signal the presence of massive bososnic fields as opposed to an exclusion. In Ref.~\cite{Fernandez:2019qbj} the authors consider the likelihood of X-ray binary BH systems stuck on Regge trajectories through a $\chi^2$ analysis for the bosonic masses required to fit the BH data.  We can define the probability that each BH currently sits on its closest Regge trajectory for a fixed boson mass, as the maximal probability a BH can deviate in the Regge plane to one of the superradiance condition boundaries (Eq.~(\ref{eq:condition})) defined by a characteristic evolutionary timescale. This defines the BH set probability that each BH sits on a regge trajectory as, 
\begin{equation}
    P_{\rm Boson}(\mathscr{M}|\{ {\rm BHs}\}) = \prod_i P_{\rm Regge}(\mathscr{M}|{\rm BH}_i)\ ,
    \label{eq:prob_fluc}
\end{equation}
where $\prod_i P_{\rm Regge}(\mathscr{M}|{\rm BH}_i)$ represents the probability the $ith$ BH sits on one of the Regge trajectories of a given model of exclusion isocontours defined by a model, $\mathscr{M}$. The value of $P_{\rm Boson}(\mathscr{M}|\{ {\rm BHs}\})$ represents the normalised cumulative probability the data set aligns to the superradiance condition boundaries and therefore the presence of a bosonic field at a particular mass. To give an indication of a possible signal in the Regge plane which might best fit current measurements we select two toy subsets of BHs from both X-ray binary sources and LIGO merger data. For X-ray binaries found in Table~\ref{tab:stellarBHs} we select the subset where all BHs possess spins, $a_* >0.5$ to enhance the likelihood each BH is susceptible to superradiance. Similarly for GW merger data we do the same by selecting BHs with spin measurements larger than the maximal uncertainty in the spin, $a_* > \pm a_*^{\rm max}$. The results of evaluating these two examples using Eq.~(\ref{eq:prob_fluc}) are presented in Fig.~\ref{fig:regge_results}. For X-ray binary data we find peaks at both $\mu_0 \simeq 4.3 \times 10^{-12}$ and at $\mu_0 \simeq 8.4 \times 10^{-12}$ the level $P_{\rm ax}(\mathcal{M}|\{ {\rm BHs}\}) \sim \mathcal{O}(10^{-8})$. For the GW merger data we find a peak at $\mu_0 \simeq 4.3 \times 10^{-12}$ at the level $P_{\rm ax}(\mathcal{M}|\{ {\rm BHs}\}) \sim \mathcal{O}(10^{-16})$. 

The low values of $P_{\rm Boson}(\mathscr{M}|\{ {\rm BHs}\})$ are indicative of the large systematics in BH parameter measurements whilst also not accounting for errors in analytical approximations used etc. Only significantly enhanced accuracy of BH measurements for a significantly larger number of events whilst incorporating all BH measurements will provide statistical clarity in this area and offers an interesting area to conduct a far more detailed analysis in the future.

\section{Multifield Phenomenology and Toy Model Measures}
\label{app:multifields}
To sample a spectrum of dimensionful quantities we associated their probability density functions to an operand measure determined by the limiting spectral statistics of a chosen general convariance matrix ensemble of the form, $\mathds{X}_{ij}=\mathscr{N}^{-1}\mathds{Y}_{ik}\Sigma_{\rm cov}\mathds{Y}_{kj}^{*}$, where $\Sigma_{\rm cov}$ is the population covariance matrix of the ensemble. Our choice of entries for the $\mathscr{N}\times \mathscr{P}$ matrix, $\mathds{Y}_{ik}$ fixes the nature of the spectrum of the $\mathscr{N}\times \mathscr{N}$ matrix $\mathds{X}_{ij}$. For any Borel probability measure (denoted $\xi$) defined over the real interval, $\xi \in [0,\infty)$, we can express its Lebesgue decomposition as $\xi = \xi^{\rm ac}+\xi^{\rm pm}+\xi^{\rm sc}$ \cite{2006math......3104T,2019arXiv190302326J}. The linearised terms, $\xi^{\rm ac}, \xi^{\rm pm}$ and $\xi^{\rm sc}$ represent the absolutely continuous, point mass and singular continuous components of the probability measure respectively. Our choice of matrix ensemble rejects the need to consider $\xi^{\rm sc}$. Strong correlated signals and non-maximal rank perturbation operations are associated to spiked outlier eigenvalues \cite{2011arXiv1101.5144M}, the location of which is represented by the perturbed point mass measure, $\xi^{\rm pm}$ (see Section~III C 3 of Ref.~\cite{Stott:2017hvl}. Likewise if $\beta > 1$ we find $\mathscr{N}-\mathscr{P}$ measures $\xi^{\rm pm}$ located at zero,  weighted by ($1-\beta$). If however we assume isometric covariance ($\Sigma_{\rm cov} \propto \mathds{I} $) and fix the upper bound $\beta<1$, then the limiting eigenvalue distribution is an unimodal analytic function determined by the limiting laws for matrices residing in the Wishart-laguerre ensemble. This pillar result of RMT we can assign to a spectrum of model parameters is the Mar\v{c}henko–Pastur distribution \cite{1967SbMat...1..457M},
\begin{equation}
    P_{\rm MP}(\lambda) = (1-\beta)\delta_0+\frac{1}{2\pi\beta\lambda\bar{\mu}^2}\sqrt{(\lambda_+-\lambda)(\lambda-\lambda_-)}\ ,
    \label{eq:mpeq}
\end{equation}
where $\lambda_+$ and $\lambda_-$ represent the spectral support supremum and infimum respectively. See Refs.~\cite{Stott:2017hvl,Stott:2018opm} for more detailed discussions on these models. 

\subsection{Bosonic Fields of Spin-0}
\label{app:spin0multi}
The general case of a multi-field spin-0 effective field theory is represented by a two derivative action of the form, 
\begin{equation}
    \mathcal{L}^{\rm ALP}_{\rm eff} = - \mathcal{K}_{ij}(\theta)\partial_{\mu}\theta^i \partial^\mu\theta^j - V(\theta^i)\ ,
    \label{eq:decayeq}
\end{equation}
defining the field basis formed by the orthogonal periods of the discrete shift symmetries of the $\mathscr{N}$ scalar fields \cite{Bachlechner:2017hsj,Bachlechner:2017zpb}. The term $\mathcal{K}_{ij}  = f_i^af_{ja}$ represents a positive-definite field space metric, whose eigenvalues determine the decay constants $f_a$, of the theory. A general multi-field potential takes the form, $V(\theta^i) = \sum^{\mathscr{N}_{\rm inst}}_{i=1}\sum^{\mathscr{N}_{\rm ax}}_{j=1}\Lambda_{i}U_i(\mathcal{Q}_{j,i}\theta_j+\delta_i)$, where $U$ is a general
periodic instanton potential with charge matrix $\mathcal{Q}$ and
phases, $\delta$. When expanding this expression about its minimum and performing a canonical normalisation ($\phi^j \equiv f^j_{a,i}\theta^i$) through a field space redefinition, we define the mass eigenstate sample basis,
\begin{equation}
    \mathcal{L}^{\rm ALP}_{\rm eff} = - \frac{1}{2}\mathcal{\delta}_{ij}\partial_{\mu}\phi^i \partial^\mu\phi^j - \frac{1}{2}\phi^{i} \mathcal{M}_{ij} \phi^{j}\ .
    \label{eq:masseq}
\end{equation}

For a simple parametric axion field theory, our choice to either sample both of the axions dimensionful quantities (Eq.~(\ref{eq:decayeq})) or just the field masses appearing in the counter term of (Eq.~(\ref{eq:masseq})), defines the number of free hyperparemters of our statistical RMT model. Ensuring that $\beta_{\mathcal{M}}\leq1$ determines no $\xi^{\rm pm}$ at zero in our limiting probability measure, loosely associated to the physical understanding each field in the spectrum acquires a mass generated from instanton effects ($\mathscr{P} \geq \mathscr{N} \equiv \mathscr{N}_{\rm inst}\geq \mathscr{N}_{\rm ax}$). For example beginning in the basis in Eq.~(\ref{eq:decayeq} defines the simple model hypervector, $\zeta = [\bar{\mu}_{\mathcal{K}},\beta_{\mathcal{K}},\bar{\mu}_{\mathcal{M}},\beta_{\mathcal{M}}]$, representing a series of terms which fix the spectral mean ($\bar{\mu}$) and extrema ($\beta$) for both the decay constants and masses. This is simply $\zeta = [\bar{\mu}_{\mathcal{M}},\beta_{\mathcal{M}}]$ if we begin in the basis of Eq.~(\ref{eq:masseq}) as we do for the examples in Section~\ref{sec:multifield}. In this case the absolvement of the decay constants moving to the mass eigenstate basis (Eq.~(\ref{eq:masseq})) acts as a pertrubative matrix operation, where free-probability ensures the traceability of the limiting spectrum of the masses. If both the kinetic matrix and mass matrix are initially constructed as \emph{white}-Wishart matrices ($\Sigma_{\rm cov} = \mathbb{I}$ and Gaussian matrix entries with zero mean etc.) then the limiting mass spectrum takes the form of the operand measure of a matrix distributed according to a scaled matrix-$F$ function \cite{bai2004,zheng2012,PAUL20141,mulder2018,2016arXiv160604417H}. See Ref.~\cite{doi:10.1137/0516047} for the limiting eigenvalue distribution of this ensemble. 

\subsection{Bosonic Fields of Spin-1}
\label{app:spin1multi}
For fields of spin-1 the simplest multi-field generalisation of Proca theory concerns a repeated set of Abelian global symmetries, $U(1)^{\mathscr{N}_{\rm Vec}}$. At the linear level of the theory the interacting case of massive spin-1 fields is understood via the canonical Proca action with an argument of the form $\mathcal{L}^{\rm 4D}_{\rm Vec}=\mathcal{L}^{\mathscr{N}_{\rm Vec}}_{\rm kin}-\mathcal{L}^{\mathscr{N}_{\rm Vec}}_{\rm Int}$, which in the quadratic limit takes form,
\begin{equation}
    \mathcal{L}^{\rm Vec}_{\rm eff} =  -\frac{1}{4}\sum^{\mathscr{N}_{\rm Vec}}_{i=1}F_{(i),\mu\nu}F^{(i),\mu\nu} - \frac{1}{2} \sum_{\alpha,\beta}A_{\mu}^{\alpha}\mathcal{M}_{\alpha\beta}A^{\beta}_{\nu}\eta^{\mu\nu}\ ,
    \label{eq:procamulti}
\end{equation}
where $F_{\mu\nu}^{i}=\partial_{\mu}A^\alpha - \partial_{\nu}A_{\mu}^\alpha$ are Maxwell tensors running over the indices $\alpha = 1,\dots,\mathscr{N}_{\rm Vec}$. Aside from the respective gauge invariances, the fields are globally invariant under concruence rotations in field space, $A_{\mu}^\alpha \rightarrow \bar{A}^\alpha_{\mu} = \mathcal{U}^\alpha_\beta A_{\mu}^\beta$. 
This ensures the freedom to work in a basis where we can trivially define the mass eigenstates of the gauge fields via the rotational invariance of $\mathcal{L}^{\mathscr{N}_{\rm Vec}}_{\rm kin}$ in Eq.~(\ref{eq:procamulti}). Assuming the rank of $\mathcal{M}_{\alpha\beta}$ is maximal in its eigendecomposition basis (i.e. all directions in field space interact through the respective mass terms) the theory is then defined with 3$\mathscr{N}_{\rm vec}$ polarisations in the four-spacetime dimensions of the effective field theory. When  $\mathcal{M}_{\alpha\beta}$ is singular with multiplicity of order the magnitude of its rank deficiency, $r$, the theory contains $\mathscr{N}_{\rm vec}-r$ massless spin-2 fields from the independent $U(1)^{\mathscr{N}_{\rm vec}-r}$ symmetries which remain. Much like the scalar case of massive axions entering the particle spectrum we can ensure that $\mathcal{M}_{\alpha\beta}$ is non-singular through the values $\beta_{\mathcal{M}}\leq 1$ in order to model a simple picture of the spectrum of $\mathscr{N}_{\rm Vec}$ spin-1 fields. using the spectral parameter $\beta_{\mathcal{M}}$ in this example to regulate the nature of the spectrum, a number of point mass measures appear at zero-bosonic mass when $\beta_{\mathcal{M}}>1$. 
\\
\section{Total Bounds Results}
Here we present the extended bounds from the results given in Section~\ref{sec:bounds} for each BH hole found in Appendix~\ref{app:bhdata}. For the free-field limit bounds (Section~\ref{sec:freefielddomain}) we calculate the exclusion regions utilising the three instability timescales discussed in Section~\ref{sec:superradiance}, $\tau_{\rm BH}\equiv \tau_{\rm SEdd} = 4.5\times 10^{6}$ yrs, $\tau_{\rm BH}\equiv \tau_{\rm Sal} = 4.5\times 10^{7}$ yrs and $\tau_{\rm BH}\equiv \tau_{\rm Hub} = 1 \times 10^{10}$ yrs. For the interacting limit we quote the self-interaction limits for each BH in Appendix~\ref{app:bhdata} 

\subsection{Free-field domain extended results}
\label{app:extendedfree}

The extended free-field domain bounds for spin-0, -1 and -2 fields for each BH are:

\onecolumngrid 
\begin{tabularx}{\textwidth}{j|s|c|c|c}
\multicolumn{5}{p{\dimexpr\textwidth-\tabcolsep\relax}}{Table VII: Bosonic field mass bounds from the stellar mass X-ray sources defined in Table~\ref{tab:stellarBHs} quoted at the 68\% confidence limit.  calculated using the analytical instability rates detailed in Section~\ref{sec:spin0fields} (spin-0), Section~\ref{sec:spin1fields} (spin-1) and Section~\ref{sec:spin2fields} (spin-2). For each BH we calculate the exclusion bounds using the timescales $\tau_{\rm SEdd}, \tau_{\rm Sal}$ and $\tau_{\rm Hub}$, denoted by the E, S and H labels respectfully. Entries containing `$-$' represent configurations where the bounds from superradiance were below the 68\% confidence interval.}\\ \multicolumn{5}{r}{}   \\
 \hline
 \noalign{\vskip 0.1cm}
  \hline
\textbf{Black Hole} & $\tau_{\rm SR}$ & \textbf{Spin-0} & \textbf{Spin-1} & \textbf{Spin-2}\\
 \hline
 \noalign{\vskip 0.1cm}
\endfirsthead
 \hline
 \noalign{\vskip 0.1cm}
  \hline
\textbf{Black Hole} & $\tau_{\rm SR}$ & \textbf{Spin-0} & \textbf{Spin-1} & \textbf{Spin-2}\\
 \hline
 \noalign{\vskip 0.1cm}
\endhead
 \hline
 \noalign{\vskip 0.1cm}
\multicolumn{5}{r}{\footnotesize( To be continued)}
\endfoot
\bottomrule
\endlastfoot
GRO J1655-40 &\tiny  \begin{tabular}{@{}c@{}c@{}}\hspace{0.005cm} E \vspace{0.1cm}\\ S\vspace{0.1cm}\\ H\end{tabular} & \scriptsize \begin{tabular}{@{}c@{}c@{}}$3.1 \times 10^{-13} {\rm eV} \leq \mu_{0} \leq 1.6 \times 10^{-11} {\rm eV}$ \\$2.3 \times 10^{-13} {\rm eV} \leq \mu_{0} \leq 1.6 \times 10^{-11} {\rm eV}$ \\$1.3 \times 10^{-13} {\rm eV} \leq \mu_{0} \leq 1.9 \times 10^{-11} {\rm eV}$\end{tabular} & \scriptsize \begin{tabular}{@{}c@{}c@{}}$4.8 \times 10^{-14} {\rm eV} \leq \mu_{1} \leq 1.6 \times 10^{-11} {\rm eV}$ \\$3.5 \times 10^{-14} {\rm eV} \leq \mu_{1} \leq 2.0 \times 10^{-11} {\rm eV}$ \\$1.8 \times 10^{-14} {\rm eV} \leq \mu_{1} \leq 2.0 \times 10^{-11} {\rm eV}$\end{tabular} & \scriptsize \begin{tabular}{@{}c@{}c@{}}$1.7 \times 10^{-13} {\rm eV} \leq \mu_{2} \leq 1.6 \times 10^{-11} {\rm eV}$ \\$1.3 \times 10^{-13} {\rm eV} \leq \mu_{2} \leq 1.6 \times 10^{-11} {\rm eV}$ \\$7.5 \times 10^{-14} {\rm eV} \leq \mu_{2} \leq 1.6 \times 10^{-11} {\rm eV}$\end{tabular} \\
\hline
\end{tabularx}
\twocolumngrid

\onecolumngrid 
\begin{tabularx}{\textwidth}{g|s|c|c|c} 
\multicolumn{5}{p{\dimexpr\textwidth-2\tabcolsep\relax}}{Table VIII: Bosonic field mass bounds from the stellar mass X-ray sources defined in Table~\ref{tab:ligodata} quoted at the 68\% confidence limit.  calculated using the analytical instability rates detailed in Section~\ref{sec:spin0fields} (spin-0), Section~\ref{sec:spin1fields} (spin-1) and Section~\ref{sec:spin2fields} (spin-2). For each BH we calculate the exclusion bounds using the timescales $\tau_{\rm SEdd}, \tau_{\rm Sal}$ and $\tau_{\rm Hub}$, denoted by the E, S and H labels respectfully. Entries containing `$-$' represent configurations where the bounds from superradiance were below the 68\% confidence interval.}\\ \multicolumn{5}{r}{}   \\
 \hline
 \noalign{\vskip 0.1cm}
  \hline
\textbf{Black Hole} & $\tau_{\rm SR}$ & \textbf{Spin-0} & \textbf{Spin-1} & \textbf{Spin-2}\\
 \hline
 \noalign{\vskip 0.1cm}
 \hline
\endfirsthead
 \hline
 \noalign{\vskip 0.1cm}
  \hline
\textbf{Black Hole} & $\tau_{\rm SR}$ & \textbf{Spin-0} & \textbf{Spin-1} & \textbf{Spin-2}\\
 \hline
 \noalign{\vskip 0.1cm}
  \hline
\endhead
 \hline
 \noalign{\vskip 0.1cm}
\multicolumn{5}{r}{\footnotesize( To be continued)}
\endfoot
\bottomrule
\endlastfoot
GW150914 (1) &\tiny  \begin{tabular}{@{}c@{}c@{}}\hspace{0.005cm} E \vspace{0.1cm}\\ S\vspace{0.1cm}\\ H\end{tabular} & \fontsize{6.5}{7} \begin{tabular}{@{}c@{}c@{}}$-$ \\$-$ \\$-$\end{tabular} & \scriptsize \begin{tabular}{@{}c@{}c@{}}$-$ \\$-$ \\$7.3 \times 10^{-15} {\rm eV} \leq \mu_{1} \leq 1.3 \times 10^{-14} {\rm eV}$\end{tabular} & \scriptsize \begin{tabular}{@{}c@{}c@{}}$-$ \\$-$ \\$-$\end{tabular} \\ \hline
 \noalign{\vskip 0.1cm}
GW150914 (2)  &\tiny  \begin{tabular}{@{}c@{}c@{}}E\\ S\\ H\end{tabular} & \scriptsize \begin{tabular}{@{}c@{}c@{}}$-$ \\$8.4 \times 10^{-14} {\rm eV} \leq \mu_{0} \leq 9.7 \times 10^{-14} {\rm eV}$ \\$4.1 \times 10^{-14} {\rm eV} \leq \mu_{0} \leq 1.0 \times 10^{-13} {\rm eV}$\end{tabular} & \scriptsize \begin{tabular}{@{}c@{}c@{}}$1.6 \times 10^{-14} {\rm eV} \leq \mu_{1} \leq 1.0 \times 10^{-13} {\rm eV}$ \\$1.1 \times 10^{-14} {\rm eV} \leq \mu_{1} \leq 1.0 \times 10^{-13} {\rm eV}$ \\$5.9 \times 10^{-15} {\rm eV} \leq \mu_{1} \leq 2.1 \times 10^{-13} {\rm eV}$\end{tabular} & \scriptsize \begin{tabular}{@{}c@{}c@{}}$5.8 \times 10^{-14} {\rm eV} \leq \mu_{2} \leq 2.1 \times 10^{-13} {\rm eV}$ \\$4.5 \times 10^{-14} {\rm eV} \leq \mu_{2} \leq 2.1 \times 10^{-13} {\rm eV}$ \\$2.5 \times 10^{-14} {\rm eV} \leq \mu_{2} \leq 3.0 \times 10^{-13} {\rm eV}$\end{tabular} \\
\hline
\end{tabularx}
\twocolumngrid

\onecolumngrid 
\begin{tabularx}{1.00\textwidth}{g|s|c|c|c}
\multicolumn{5}{p{\dimexpr\textwidth-2\tabcolsep\relax}}{Table IX: Bosonic field mass bounds from the stellar mass X-ray sources defined in Table~\ref{tab:SMBH} quoted at the 68\% confidence limit.  calculated using the analytical instability rates detailed in Section~\ref{sec:spin0fields} (spin-0), Section~\ref{sec:spin1fields} (spin-1) and Section~\ref{sec:spin2fields} (spin-2). For each BH we calculate the exclusion bounds using the timescales $\tau_{\rm SEdd}, \tau_{\rm Sal}$ and $\tau_{\rm Hub}$, denoted by the E, S and H labels respectfully. Entries containing `$-$' represent configurations where the bounds from superradiance were below the 68\% confidence interval.}\\ \multicolumn{5}{r}{}   \\
 \hline
 \noalign{\vskip 0.1cm}
 \hline
\textbf{Black Hole} & $\tau_{\rm SR}$ & \textbf{Spin-0} & \textbf{Spin-1} & \textbf{Spin-2}\\
 \hline
 \noalign{\vskip 0.1cm}
 \hline
\endfirsthead
 \hline
 \noalign{\vskip 0.1cm}
  \hline
\textbf{Black Hole} & $\tau_{\rm SR}$ & \textbf{Spin-0} & \textbf{Spin-1} & \textbf{Spin-2}\\
 \hline
 \noalign{\vskip 0.1cm}
 \hline
\endhead
\multicolumn{5}{r}{\footnotesize( To be continued)}
\endfoot
\bottomrule
\endlastfoot
Fairall 9 &\tiny  \begin{tabular}{@{}c@{}c@{}}\hspace{0.005cm} E \vspace{0.1cm}\\ S\vspace{0.1cm}\\ H\end{tabular} & \scriptsize \begin{tabular}{@{}c@{}c@{}}$-$ \\$-$ \\$2.5 \times 10^{-20} {\rm eV} \leq \mu_{0} \leq 5.9 \times 10^{-20} {\rm eV}$\end{tabular} & \scriptsize \begin{tabular}{@{}c@{}c@{}}$1.6 \times 10^{-20} {\rm eV} \leq \mu_{1} \leq 6.1 \times 10^{-20} {\rm eV}$ \\$1.1 \times 10^{-20} {\rm eV} \leq \mu_{1} \leq 6.1 \times 10^{-20} {\rm eV}$ \\$5.0 \times 10^{-21} {\rm eV} \leq \mu_{1} \leq 1.2 \times 10^{-19} {\rm eV}$\end{tabular} & \scriptsize \begin{tabular}{@{}c@{}c@{}}$3.7 \times 10^{-20} {\rm eV} \leq \mu_{2} \leq 1.2 \times 10^{-19} {\rm eV}$ \\$2.8 \times 10^{-20} {\rm eV} \leq \mu_{2} \leq 1.2 \times 10^{-19} {\rm eV}$ \\$1.5 \times 10^{-20} {\rm eV} \leq \mu_{2} \leq 1.8 \times 10^{-19} {\rm eV}$\end{tabular} \\ \hline
 \noalign{\vskip 0.1cm}  Mrk 79  &\tiny  \begin{tabular}{@{}c@{}c@{}}\hspace{0.005cm} E \vspace{0.1cm}\\ S\vspace{0.1cm}\\ H\end{tabular} & \scriptsize \begin{tabular}{@{}c@{}c@{}}$2.5 \times 10^{-19} {\rm eV} \leq \mu_{0} \leq 4.6 \times 10^{-19} {\rm eV}$ \\$1.8 \times 10^{-19} {\rm eV} \leq \mu_{0} \leq 8.8 \times 10^{-19} {\rm eV}$ \\$1.0 \times 10^{-19} {\rm eV} \leq \mu_{0} \leq 1.3 \times 10^{-18} {\rm eV}$\end{tabular} & \scriptsize \begin{tabular}{@{}c@{}c@{}}$5.7 \times 10^{-20} {\rm eV} \leq \mu_{1} \leq 9.1 \times 10^{-19} {\rm eV}$ \\$4.3 \times 10^{-20} {\rm eV} \leq \mu_{1} \leq 9.1 \times 10^{-19} {\rm eV}$ \\$2.0 \times 10^{-20} {\rm eV} \leq \mu_{1} \leq 1.4 \times 10^{-18} {\rm eV}$\end{tabular} & \scriptsize \begin{tabular}{@{}c@{}c@{}}$1.4 \times 10^{-19} {\rm eV} \leq \mu_{2} \leq 1.4 \times 10^{-18} {\rm eV}$ \\$1.1 \times 10^{-19} {\rm eV} \leq \mu_{2} \leq 1.4 \times 10^{-18} {\rm eV}$ \\$5.7 \times 10^{-20} {\rm eV} \leq \mu_{2} \leq 1.8 \times 10^{-18} {\rm eV}$\end{tabular} \\
\hline
\end{tabularx}
\twocolumngrid
 \subsection{Interacting domain extended results}
\label{app:extendedinteracting}

The extended interacting domain bounds for each BH are:
 
\onecolumngrid 
\begin{tabularx}{1.00\textwidth}{b|c|b|c|b|c}
\multicolumn{6}{p{\dimexpr\textwidth-2\tabcolsep\relax}}{Table X: Bounds on axion decay constants $f_a$ quoted at the 68\% confidence interval determined using the method discussed in Section~\ref{sec:interactingresults}. Each constraint for the BHs detailed in Appendix~\ref{app:bhdata} are determined using Eq.~\ref{eq:boserate}. Entries containing `$-$' represent configurations where the bounds from superradiance were below the 68\% confidence interval.}\\ \multicolumn{5}{r}{}   \\ 
\hline
\hline
\textbf{Black Hole} & \textbf{Bound} & \textbf{Black Hole} & \textbf{Bound}& \textbf{Black Hole} & \textbf{Bound}\\
\hline
 \noalign{\vskip 0.1cm}
 \hline
\endfirsthead
\hline
 \noalign{\vskip 0.1cm}
 \hline
\textbf{Black Hole} &  \textbf{Bound} & \textbf{Black Hole} & \textbf{Bound}& \textbf{Black Hole} & \textbf{Bound}\\
\hline
 \noalign{\vskip 0.1cm}
 \hline
\endhead
\multicolumn{6}{r}{\footnotesize( To be continued)}
\endfoot
\bottomrule
 \noalign{\vskip 0.1cm} 
\endlastfoot
 \noalign{\vskip 0.1cm} 
GRO J1655-40 &$f_a \geq 5.1 \times 10^{14}\ {\rm GeV}$ & GRS 1716-249 &$f_a \geq 1.8 \times 10^{14}\ {\rm GeV}$  & A 0620-00  & -  \\
\hline
 \noalign{\vskip 0.1cm} 
LMC X-3 &$f_a \geq 4.9 \times 10^{13}\ {\rm GeV}$ & XTE J1550-564  &$f_a \geq 5.4 \times 10^{13}\ {\rm GeV}$ & 4U 1543-475 & $f_a \geq 2.6 \times 10^{13}\ {\rm GeV}$  \\
\hline
 \noalign{\vskip 0.1cm} 
 LMC X-1 &$f_a \geq 5.5 \times 10^{14}\ {\rm GeV}$ & GRS 1915+105 & $f_a \geq 2.5 \times 10^{14}\ {\rm GeV}$& GRS 1124-683 & $f_a \geq 8.2 \times 10^{13}\ {\rm GeV}$  \\
\hline
 \noalign{\vskip 0.1cm} 
Cygnus X-1 &$f_a \geq 3.8 \times 10^{14}\ {\rm GeV}$& M33 X-7 &$f_a \geq 1.5 \times 10^{15}\ {\rm GeV}$ &\\
\hline
 \noalign{\vskip 0.1cm} 
 \hline
GW150914 (1) &-& GW150914 (2) &- &GW151012 (1)&-\\
\hline
 \noalign{\vskip 0.1cm} 
GW151012 (2) &$f_a \geq 5.4 \times 10^{13}\ {\rm GeV}$& GW151226 (1) &$f_a \geq 8.4 \times 10^{13}\ {\rm GeV}$ &GW151226 (2)&$f_a \geq 1.4 \times 10^{14}\ {\rm GeV}$\\
\hline
 \noalign{\vskip 0.1cm} 
GW170104 (1)&-& GW170104 (2)  &$f_a \geq 4.9 \times 10^{13}\ {\rm GeV}$ &GW170608 (1)& - \\
\hline
 \noalign{\vskip 0.1cm} 
GW170608 (2) &$f_a \geq 5.1 \times 10^{13}\ {\rm GeV}$& GW170729 (1) &$f_a \geq 4.4 \times 10^{13}\ {\rm GeV}$&GW170729 (2) &$f_a \geq 2.3 \times 10^{14}\ {\rm GeV}$\\
\hline
\end{tabularx}
\twocolumngrid

\bibliography{apssamp}

\end{document}